\documentclass[aps,prd,superscriptaddress]{revtex4}
\usepackage{epsfig,epsf}
\usepackage{amsmath}
\usepackage{amsthm}
\usepackage{amsfonts}
\usepackage{amssymb}
\usepackage{dsfont}
\usepackage{multirow}
\usepackage{appendix}
\usepackage{slashed}
\usepackage[active]{srcltx}
\usepackage{psfrag}
\usepackage{multirow}
\begin{document}

\title{Structure of the  $\Xi_b(6227)^-$ Resonance}
\date{\today}
\author{T.~M.~Aliev}
\affiliation{Physics Department,
Middle East Technical University, 06531 Ankara, Turkey}
\author{K.~Azizi}
\affiliation{Physics Department, Do\u gu\c s University,
Ac{\i}badem-Kad{\i}k\"oy, 34722 Istanbul, Turkey}
\affiliation{School of Physics, Institute for Research in Fundamental Sciences (IPM), P. O. Box 19395-5531, Tehran, Iran}
\author{Y.~Sarac}
\affiliation{Electrical and Electronics Engineering Department,
Atilim University, 06836 Ankara, Turkey}
\author{H.~Sundu}
\affiliation{Department of Physics, Kocaeli University, 41380 Izmit, Turkey}

\begin{abstract}
We explore the recently observed $\Xi_b(6227)^-$ resonance to fix  its quantum numbers. To this end, we consider various possible scenarios: It can be considered as either $1P$/$2S$ excitations of the  $\Xi_b^-$ and $\Xi_b'(5935)^−$ ground state baryons with spin-$\frac{1}{2}$ or  $1P$/$2S$ excitations of the ground-state $\Xi_b(5955)^−$ with spin-$\frac{3}{2}$. We calculate the masses of the possible angular-orbital $1P$ and $2S$ excited states corresponding to each channel employing the  QCD sum rule technique. It is seen that all the obtained masses are in agreement with the experimentally observed value, implying that the mass calculations are not enough to determine the quantum numbers of the state under question. Therefore, we extend the analysis to investigate the possible decays of the  excited states into $\Lambda_b^0 K^-$ and $\Xi_b^-\pi$. Using the light cone QCD sum rule method, we calculate the corresponding strong coupling constants, which are used to extract the  decay widths of the modes under consideration. Our results on decay widths indicate that the $\Xi_b(6227)^-$ is $1P$ angular-orbital excited state of the $\Xi_b(5955)^−$ baryon with quantum numbers $J^P=\frac{3}{2}^-$.

\end{abstract}

\maketitle
\section{Introduction}
The theoretical studies of the heavy baryons involving their spectroscopic parameters and the interaction mechanisms improve our understanding on the nonperturbative regime of the strong interaction, as well as their nature and internal structure. As a result of the impressive developments in the experimental sector in the last decade, almost all of the ground state baryons with single heavy quark were observed \cite{Tanabashi2018,Basile81,Abreu95,Aaltonen:2007ar,Abazov:2008qm,Aaij:2012da,Chatrchyan:2012ni,Aaltonen:2013tta,Aaij:2014yka}.  

The spectroscopy of the heavy baryons containing $b$ quark has been investigated within different models. These approaches include the quark model~\cite{Capstick:1986bm,Ebert:2005xj,Ebert:2007nw,Garcilazo:2007eh,Karliner:2007jp,Karliner:2007cu,Roberts:2007ni,Karliner:2008sv,Ebert:2011kk,Yoshida:2015tia}, the QCD sum rule approach~\cite{Liu:2007fg,Wang:2010vn,Wang:2010it,Mao:2015gya,Agaev17epl}, lattice QCD~\cite{Lewis:2008fu}, $1/N_c $ and $1/m_b$ expansions \cite{E.Jenkins} and Faddeev approach~\cite{Valcarce:2008dr}. To gain deeper understanding on the single bottom baryons, their properties such as magnetic dipole moments and strong decays were  investigated in Refs.~\cite{Aliev:2014jra,Aliev:2014bma,Aliev:2015xaa,Aliev:2015wba,Agamaliev:2016fou,Aliev:2017tej}. In Ref.~\cite{Wang:2017kfr}, their strong and radiative decays were studied using the constituent quark model.

The quark model predicts existence of many new baryons with one, two, or three heavy quarks. The impressive developments in the experimental techniques indicate that more heavy baryons would be observed in  near future. With this motivation and the motivation brought by the recent observation of the LHCb Collaboration~\cite{Aaij:2018yqz}, we investigate the masses and decay constants of the low-lying $ 2S $ and $ 1P $ excited $\Xi_b$ baryons having $J=\frac{1}{2}$ and $J=\frac{3}{2}$. The LHCb collaboration  recently reported the observation of $\Xi_b(6227)^-$ with mass $m_{\Xi_b(6227)^-}=6226.9\pm 2.0\pm 0.3\pm 0.2$~MeV and width $\Gamma_{\Xi_b(6227)^-}=18.1\pm5.4\pm 1.8$~MeV. From the observed mass and decay modes, it was stated that $\Xi_b(6227)^-$ state may be $1P$ or $2S$ excited baryon (see also~\cite{Aaij:2018yqz}). After this observation, this state is considered in Ref.~\cite{Chen:2018orb} and its mass and strong decays were analyzed.  The obtained results indicated the possibility of its being a $P$-wave state with $J^P=\frac{3}{2}^-$ or $\frac{5}{2}^-$. It is clear that to identify the characteristics of this baryon, more studies are necessary. In this study, we aim to identify the properties of the $\Xi_b(6227)^-$ state by using  QCD sum rule method \cite{Shifman:1978bx,Shifman:1978by,Ioffe81}. This method serves as one of the powerful nonperturbative methods. In the calculations, the observed state is considered as $ 1P $ and $ 2S $ excitations of the ground state $\Xi_b$ baryons having $J=\frac{1}{2}$ and $J=\frac{3}{2}$ and $\Xi_b^{\prime}$ baryon having $J=\frac{1}{2}$. We obtain the decay constants and masses of the $ 1P $ and $ 2S $ excitations for each case. In the quark model's notations, these states are represented by $1^2 P_{1/2}$, $2^2 S_{1/2}$, $1^4 P_{3/2}$, and $2^4 S_{3/2}$. For simplicity,  we  denote these states  as $J^P=1/2^- $, $J^P=1/2^+$, $J^P=3/2^- $, and  $J^P=3/2^+$, respectively.

Moreover, we discuss the strong transitions of the $\Xi_b(6227)^-$ to $\Lambda_b^0$ and $K^-$ as well as  to $\Xi_b^0$ and $\pi^-$. To this end we use an extension of the traditional QCD sum rules method, namely the light cone QCD sum rule (LCSR) technique~\cite{Braun:1988qv, Balitsky:1989ry,Chernyak:1990ag}. In the LCSR, the time-ordered product of the interpolating currents is sandwiched between an on-shell state and the vacuum. The on-shell state in the present work is either $K$ or $\pi$ meson depending on the considered transition. Again taking into account the $ 1P $ and $ 2S $ excitation possibilities for $\Xi_b(6227)^-$ state, the related coupling constants and decay widths are obtained for the mentioned transitions. The results are compared to the existing experimental ones with the aim of better determination of the quantum numbers to be assigned to the $\Xi_b(6227)^-$ state.

The outline of the article is as follows: In Section II, we derive the QCD sum rules for the masses and decay constants of the $\Xi_b(6227)^-$ with the possible quantum  numbers $J^P=\frac{1}{2}^{\pm}$  and $\frac{3}{2}^{\pm}$. This section also contains the numerical values obtained for masses and decay constants of considered states which will be used as inputs in the following section. In Section III, within light cone QCD sum rule method, we obtain the coupling constants for the considered transitions with possible configurations assigned to the $\Xi_b(6227)^-$ and present the results obtained from the analysis. This section includes also the decay width calculations for the considered transitions. Section IV is devoted to the summary and discussion of the results.

\section{Spectroscopic Parameters of the $\Xi_b$ States}

In this section, the details of the calculations for spectroscopic properties, i.e. masses and decay constants, of $ 1P $ and $ 2S $ excited $\Xi_b$ states are presented for three different total angular momentum $J$ possibilities. The calculations for all three considerations are performed using the QCD sum rule formalism which starts from the following correlation function
\begin{equation}
\Pi _{(\mu \nu)}(q)=i\int d^{4}xe^{iq\cdot x}\langle 0|\mathcal{T}\{J_{B(\mu)}(x){J^{\dagger}_{B(\nu)}}(0)\}|0\rangle ,  \label{eq:CorrF1}
\end{equation}        
The correlation function is written in terms of the interpolating current of the considered state, i.e. the current $J_{B(\mu)}$ corresponding to the considered $J=\frac{1}{2}(\frac{3}{2})$ state, which is formed using quark fields and considering the quantum numbers of the state. The sub-index $B$ represents one of the states, $\Xi_b$ ($J=\frac{1}{2}$), $\Xi_b^{'}$  ($J=\frac{1}{2}$) or $\Xi_b$  ($J=\frac{3}{2}$). We will use the following interpolating currents in the calculations:

$a)$ For $J= \frac{1}{2} $ particles: 
\begin{eqnarray}\label{Eq:Current1a}
\label{eapo03}
J_{\Xi_b} &=& {1\over \sqrt{6}} \epsilon^{abc} \Big\{ 2 (d^{aT} C
s^b) \gamma_5 b^c + (d^{aT} C b^b) \gamma_5 s^c +
(b^{aT} C s^b) \gamma_5 d^c + 2 \beta (d^{aT} C \gamma_5 s^b) b^c \nonumber\\&+& 
\beta (d^{aT} C \gamma_5 b^b) s^c +\beta
(b^{aT} C \gamma_5 s^b) d^c \Big\}~, \nonumber \\
J_{\Xi_b^{\prime}} &=& -{1\over \sqrt{2}} \epsilon^{abc} \Big\{ (d^{aT} C
b^b) \gamma_5 s^c - (b^{aT} C s^b) \gamma_5 d^c + \beta
(d^{aT} C \gamma_5 b^b) s^c - \beta (b^{aT} C \gamma_5 s^b) d^c
\Big\}.
\end{eqnarray}
$b)$ For $J=\frac{3}{2} $ particles:
\begin{eqnarray}
\label{Eq:Current1b}
J_{{\Xi_b},{\mu}} = \sqrt{\frac{2}{3}} \epsilon^{abc} \Big\{ (d^a C \gamma_\mu s^b) b^c + (s^a C
\gamma_\mu b^b) d^c + (b^a C \gamma_\mu d^b) s^c \Big\}~.
\end{eqnarray}
The indices $a$, $b$, and $c$ in the current expressions are used to represent the color indices, $C$ is the charge conjugation operator, and $\beta$ present in the $ J=\frac{1}{2}$ currents is an arbitrary parameter.   

In QCD sum rule calculations, we calculate the correlator in two ways. In the first step, it is calculated in terms of hadronic degrees of freedom, considering the interpolating fields as operators annihilating or creating those hadrons. This side is expressed in terms of hadronic degrees of freedom and denoted as physical or phenomenological side. For the calculation of this side,  complete sets of hadronic states with the same quantum numbers of the considered hadrons are inserted in the correlation function. As a result we have
\begin{eqnarray}
\Pi_{(\mu\nu)}^{\mathrm{Phys}}(q)&=&\frac{\langle 0|J_{B(\mu )} |B(q,s)\rangle \langle B(q,s)|\bar{J}_{B(\nu)}|0\rangle}{m^{2}-q^{2}}
+\frac{\langle 0|J_{B(\mu )} |\widetilde{B}(q,s)\rangle \langle \widetilde{B}(q,s)|\bar{J}_{B(\nu)}|0\rangle}{\widetilde{m}^{2}-q^{2}}
+\ldots,
\label{eq:phys1a}
\end{eqnarray}
and
\begin{eqnarray}
\Pi_{(\mu\nu)}^{\mathrm{Phys}}(q)&=&\frac{\langle 0|J_{B(\mu) } |B(q,s)\rangle \langle B(q,s)|\bar{J}_{B(\nu)}|0\rangle}{m^{2}-q^{2}}
+\frac{\langle 0|J_{B(\mu) } |B'(q,s)\rangle \langle B'(q,s)|\bar{J}_{B(\nu)}|0\rangle}{m'^2-q^{2}}
+\ldots,
\label{eq:phys1b}
\end{eqnarray}
when the $ 1P $  and the $ 2S $ excitations are considered, respectively. Here $m$, $\widetilde{m}$ and $m'$ are the mass of the ground, $ 1P $ and $ 2S $ excited states of the $\Xi_b$ baryons, correspondingly. $J_{B(\mu)}$ represents either the current $J_B$ of $J=\frac{1}{2}$ or that $J_{B\mu}$ of $J=\frac{3}{2}$ baryon. The contributions of higher states and the continuum are represented by the dots. The matrix elements between the vacuum and one-particle states are defined as 
\begin{eqnarray}
\langle 0|J_B |B(q,s)\rangle &=&\lambda u(q,s),
\nonumber \\
\langle 0|J_B |\widetilde{B}(q,s)\rangle
 &=&\widetilde{\lambda}\gamma_5 u(q,s),
\nonumber \\
\langle 0|J_B |B'(q,s)\rangle
 &=&\lambda' u(q,s),
\label{eq:Res1}
\end{eqnarray}
%\
for $J=\frac{1}{2}$ states and
\begin{eqnarray}
\langle 0|J_{B\mu } |B^{*}(q,s)\rangle &=&\lambda^{*}u_{\mu}(q,s),
\nonumber \\
\langle 0|J_{B\mu } |\widetilde{B}^{*}(q,s)\rangle
 &=&\widetilde{\lambda}^{*}\gamma_5u_{\mu}(q,s),
\nonumber \\
\langle 0|J_{B\mu } |B^{*}{}'(q,s)\rangle
 &=&\lambda^{*}{}'u_{\mu}(q,s),
\label{eq:Res2ek}
\end{eqnarray}
%\
for $J=\frac{3}{2}$ states. In this section and in the following one, $B(B^{*})$, $\widetilde{B}(\widetilde{B}^{*})$ and $B'(B^{*}{}')$ notations are used to represent the ground $ 1S $, $ 1P $ and $ 2S $ excited states corresponding to the $J=\frac{1}{2}(\frac{3}{2})$ baryon, respectively and $\lambda(\lambda^{*})$,  $\widetilde{\lambda}(\widetilde{\lambda}^{*})$ and $\lambda'(\lambda^{*}{}')$ are the decay constants related to each of these states. In Eq. (\ref{eq:Res2ek})  $u_{\mu}(q,s)$  is the Rarita-Schwinger spinor for the $J=\frac{3}{2}$ states. Summation over the spins of  spinors is performed by using the formulas
\begin{eqnarray}\label{Dirac}
\sum_s  u (q,s)  \bar{u} (q,s) &= &(\!\not\!{q} + m),
\end{eqnarray}
and
\begin{eqnarray}\label{Rarita}
\sum_s  u_{\mu} (q,s)  \bar{u}_{\nu} (q,s) &= &-(\!\not\!{q} + m)\Big[g_{\mu\nu} -\frac{1}{3} \gamma_{\mu} \gamma_{\nu} - \frac{2q_{\mu}q_{\nu}}{3m^{2}} +\frac{q_{\mu}\gamma_{\nu}-q_{\nu}\gamma_{\mu}}{3m} \Big].
\end{eqnarray}
Using these expressions together with the matrix elements in Eqs.~(\ref{eq:phys1a}) and (\ref{eq:phys1b}), we get the following expressions  for the $J=\frac{1}{2}$ case:
\begin{eqnarray}\label{PhyssSide1a}
\Pi^{\mathrm{Phys}}(q)&=&\frac{\lambda^{2}(\!\not\!{q}+m)}{m^{2}-q^2}+\frac{{\widetilde{\lambda}}^{2}(\!\not\!{q}-\widetilde{m})}{\widetilde{m}^{2}-q^2}+\ldots,
\end{eqnarray}
and
\begin{eqnarray}\label{PhyssSide1b}
\Pi^{\mathrm{Phys}}(q)&=&\frac{\lambda^{2}(\!\not\!{q}+m)}{m^{2}-q^2}+\frac{\lambda'{}^{2}(\!\not\!{q}+m')}{m'^{2}-q^2}+\ldots.
\end{eqnarray}
For $J=\frac{3}{2}$ case, we obtain
\begin{eqnarray}\label{PhyssSide}
\Pi_{\mu\nu}^{\mathrm{Phys}}(q)&=&-\frac{\lambda^{*}{}^{2}}{q^{2}-m^{*}{}^{2}}(\!\not\!{q} + m^{*})\Big[g_{\mu\nu} -\frac{1}{3} \gamma_{\mu} \gamma_{\nu} - \frac{2q_{\mu}q_{\nu}}{3m^{*}{}^{2}} +\frac{q_{\mu}\gamma_{\nu}-q_{\nu}\gamma_{\mu}}{3m^{*}} \Big]\nonumber \\
&-&\frac{\widetilde{\lambda}^{*}{}^{2}}{q^{2}-\widetilde{m}^{*}{}^{2}}(\!\not\!{q} - \widetilde{m}^{*})\Big[g_{\mu\nu} -\frac{1}{3} \gamma_{\mu} \gamma_{\nu} - \frac{2q_{\mu}q_{\nu}}{3\widetilde{m}^{*}{}^{2}} +\frac{q_{\mu}\gamma_{\nu}-q_{\nu}\gamma_{\mu}}{3\widetilde{m}^{*}} \Big]+\ldots,
\end{eqnarray}
and
\begin{eqnarray}\label{PhyssSide}
\Pi_{\mu\nu}^{\mathrm{Phys}}(q)&=&-\frac{\lambda^{*}{}^{2}}{q^{2}-m^{*}{}^{2}}(\!\not\!{q} + m^{*})\Big[g_{\mu\nu} -\frac{1}{3} \gamma_{\mu} \gamma_{\nu} - \frac{2q_{\mu}q_{\nu}}{3m^{*}{}^2} +\frac{q_{\mu}\gamma_{\nu}-q_{\nu}\gamma_{\mu}}{3m^{*}} \Big]\nonumber\\&-&\frac{\lambda'{}^{2}}{q^{2}-m^{*}{}'{}^{2}}(\!\not\!{q} + m^{*}{}')\Big[g_{\mu\nu} -\frac{1}{3} \gamma_{\mu} \gamma_{\nu} - \frac{2q_{\mu}q_{\nu}}{3m^{*}{}'{}^{2}} +\frac{q_{\mu}\gamma_{\nu}-q_{\nu}\gamma_{\mu}}{3m^{*}{}'} \Big]+\ldots,
\end{eqnarray}
where, the first and second equations for each case belong to the $ 1P $ and $ 2S $ excitations, respectively. Note that  Eqs.~(\ref{PhyssSide1a}) and (\ref{PhyssSide1b}) are used as common formula for both $J=\frac{1}{2}$ $\Xi_b$ and $\Xi_b^{'}$ baryons.  

The same correlation function, Eq.~(\ref{eq:CorrF1}), can also be calculated in terms of quarks and gluons using the explicit expressions of the interpolating currents and by the help of operator product expansion (OPE). The possible contractions between quark fields render the expression into a form, which contains the heavy and light quark propagators. Using these quark propagators in coordinate space  and performing the necessary Fourier transformations  to transfer the calculations to the momentum space, we get the results for the QCD side. 
    
After  calculations of the physical and QCD sides, Borel transformations are applied to both sides with the aim of suppression of the higher states and continuum contributions. Finally, we choose the coefficients of the same Lorentz structures from both sides to get the QCD sum rules for the masses and decay constants. In  calculations the structures $ \!\not\!{q}$ and $I$ are chosen for $J=\frac{1}{2}$ cases and $ \!\not\!{q}g_{\mu\nu}$ and $g_{\mu\nu}$ for $J=\frac{3}{2}$ case.  In the case of $J=\frac{3}{2}$, actually there are more  Lorentz structures; however, except these ones, the others contain contributions from the $J=\frac{1}{2}$ states and, to avoid the contributions of these undesired states, only these two structures are considered. Final form of the QCD sum rules for the masses and decay constants are obtained as   
\begin{eqnarray}
\lambda^{2} e^{-\frac{m^{2}}{M^{2}}}+\widetilde{\lambda}^2(\lambda'{}^{ 2}) e^{-\frac{\widetilde{m}^2(m'{}^{2})}{M^{2}}}&=&\Pi^{\mathrm{QCD}}_{1},
\nonumber \\
m \lambda^{2} e^{-\frac{m^{2}}{M^{2}}}\mp \widetilde{m}(m') \widetilde{\lambda}^{2}(\lambda'{}^{2}) e^{-\frac{\widetilde{m}^{2}(m'{}^2)}{M^{2}}}&=&\Pi^{\mathrm{QCD}}_{2}.
\label{Eq:sumrule2}
\end{eqnarray}
where $-$ and $+$ signs in the second equation correspond to $ 1P $ excited $\widetilde{B}$ and $ 2S $ excited $B'$ states, respectively. $\Pi^{\mathrm{QCD}}_{i}$ with $i=1,2$ are the Borel transformed coefficients of the structures $ \!\not\!{q}$ and $I$ for $J=\frac{1}{2}$ cases obtained in QCD sides. The results for $J=\frac{3}{2}$ case can be obtained from Eq.~(\ref{Eq:sumrule2}) by replacing $\widetilde{\lambda} \rightarrow \widetilde{\lambda}^{*}$, $\lambda' \rightarrow \lambda^{*}{}'$, $\widetilde{m} \rightarrow \widetilde{m}^{*}$, $m' \rightarrow m^{*}{}'$, and $\Pi^{\mathrm{QCD}}_{i} \rightarrow \Pi^{*\mathrm{QCD}}_{i}$ for the coefficients of $\!\not\!{q}g_{\mu\nu}$  and $g_{\mu\nu}$ attained in QCD side.

To perform the numerical analysis, various input parameters entering to the sum rules are needed. Some of these input parameters are given in Table~\ref{tab:Param}. 
\begin{table}[tbp]
%\rowcolors{1}{lightgray}{white}
\begin{tabular}{|c|c|}
\hline\hline
Parameters & Values \\ \hline\hline
$ m_{\Xi_b^{-}}$                         & $5794.5\pm 1.4~\mathrm{MeV}$ \cite{Tanabashi2018}\\
$m_{\Xi_b^{\prime}(5935)^-}$               & $5935.02\pm0.02\pm 0.05~\mathrm{MeV}$ \cite{Tanabashi2018}\\
$ m_{\Xi_b(5955)^{-}}$                   & $5955.33\pm 0.12\pm 0.05~\mathrm{MeV}$ \cite{Tanabashi2018}\\
$m_{b}$                                  & $4.18^{+0.04}_{-0.03}~\mathrm{GeV}$ \cite{Tanabashi2018}\\
$m_{s}$                                  & $128^{+12}_{-4}~\mathrm{MeV}$ \cite{Tanabashi2018}\\
$m_{d}$                                  & $4.7^{+0.5}_{-0.3}~\mathrm{MeV}$ \cite{Tanabashi2018}\\
$ \lambda_{\Lambda_b}$                   & $(3.85\pm 0.56)\times 10^{-2}~\mathrm{GeV}^3$ \cite{Azizi:2008ui}\\
$ \lambda_{\Xi_b}$                       & $0.054\pm 0.012~\mathrm{GeV}^3$ \cite{Azizi:2016dmr}\\
$\langle \bar{q}q \rangle$ & $(-0.24\pm 0.01)^3$ $\mathrm{GeV}^3$ \cite{Belyaev:1982sa}  \\
$\langle \bar{s}s \rangle $              & $0.8\langle \bar{q}q \rangle$ \cite{Belyaev:1982sa} \\
$m_{0}^2 $                               & $(0.8\pm0.1)$ $\mathrm{GeV}^2$ \cite{Belyaev:1982sa}\\
$\langle g_s^2 G^2 \rangle $             & $4\pi^2 (0.012\pm0.004)$ $~\mathrm{GeV}
^4 $\cite{Belyaev:1982cd}\\
$ \Lambda $                              & $ (0.5\pm0.1) $ $\mathrm{GeV} $ \cite{Chetyrkin:2007vm} \\
\hline\hline
\end{tabular}%
\caption{Some input parameters used in the calculations of masses and coupling constants.}
\label{tab:Param}
\end{table}
Besides these parameters, the sum rules contain three auxiliary parameters. These are the Borel parameter $M^2$, threshold parameter $s_0$, and an arbitrary parameter $\beta$ existing in the calculations of $J=\frac{1}{2}$ states. To fix their working intervals, we follow the standard criteria of the QCD sum rules formalism. To begin with, in the determination of threshold parameter $s_0$, we need to emphasize that it is not completely arbitrary and has a relation with the energy of first excited state having the same quantum numbers with the considered state. However, since we have very limited knowledge on the energy of excited states, we fix its interval looking at the pole dominance condition. We demand that the pole contributions for each case are dominant and comprise the highest part of the total value. The Borel parameter region is determined looking at the convergence of the OPE. This requires also the dominance of the perturbative terms over the nonperturbative ones in the calculations. Claiming these, the lower limit of the Borel parameter is fixed. For the upper limit of this parameter, the pole dominance is required. Finally, for the calculations including the parameter $\beta$, its working interval is obtained from the analysis of the results, requiring least possible dependence on this parameter. For this purpose, one examines the variance of the results as a function of $\cos\theta$, where $\beta=\tan\theta$, and determines the regions where the results have relatively weak dependence on $\cos\theta$. Actually, the relatively weak dependence on the auxiliary parameters is another requirement in the QCD sum rules calculation to gain reliable results for the physical parameters under consideration. With all these requirements, the intervals for  auxiliary parameters, for all the considered states, are attained as
\begin{eqnarray}
45~\mbox{GeV}^2 \leq s_0 \leq 48~\mbox{GeV}^2
\end{eqnarray}
\begin{eqnarray}
6~\mbox{GeV}^2 \leq M^2\leq 9~\mbox{GeV}^2
\end{eqnarray}
and
\begin{eqnarray}
-1\leq\cos\theta\leq -0.3 ~~~~~\mbox{and} ~~~~~~0.3\leq \cos\theta\leq 1 
\end{eqnarray}

\begin{figure}[h!]
\begin{center}
\includegraphics[totalheight=5cm,width=7cm]{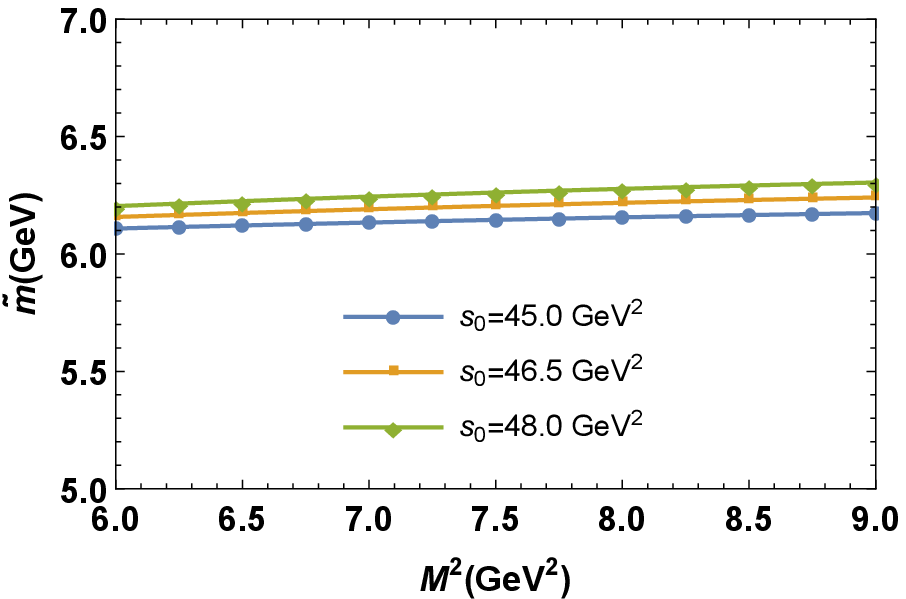}
\includegraphics[totalheight=5cm,width=7cm]{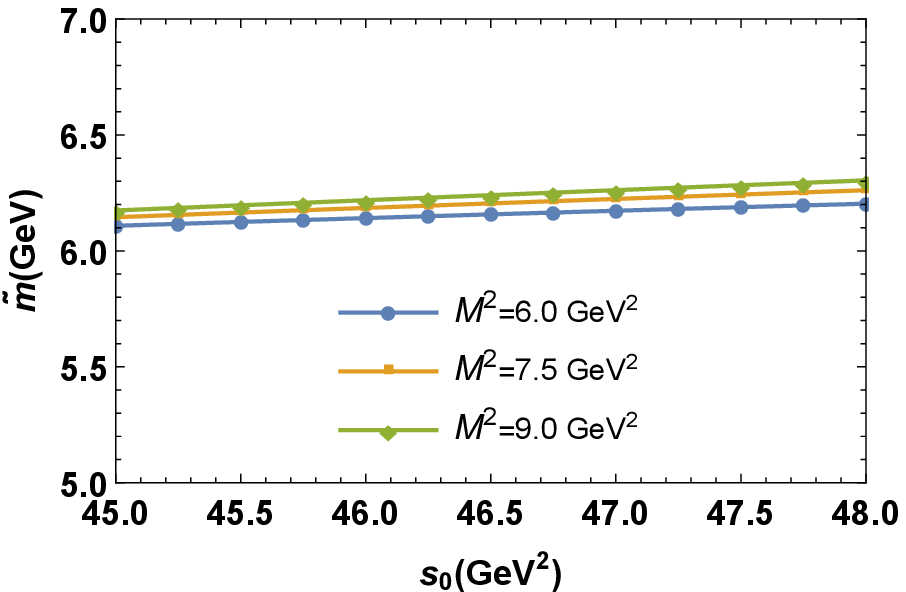}
\end{center}
\caption{\textbf{Left:} The mass $\widetilde{m}$ for the  $ 1P $ excitation of $\Xi_b(5955)^{-}$ baryon vs Borel parameter $M^2$.
\textbf{Right:} The mass $\widetilde{m}$ for the $ 1P $ excitation of $\Xi_b(5955)^{-}$ baryon  vs threshold parameter $s_0$.}
\label{gr:mXib}
\end{figure}

Using the working regions for the auxiliary parameters, together with the parameters given in Table~\ref{tab:Param}, we obtain the final results for the masses and  decay constants under consideration. Note that the ground state mass values of $\Xi_b$ baryons, i.e. the masses of $\Xi_b^-$, $\Xi_b^{\prime}(5935)^-$ and $\Xi_b(5955)^{-}$ are taken as inputs in the equations. The values of the the masses corresponding to $ 1P $ and $ 2S $ excitations are obtained as presented  in Table~\ref{tab:results}. 
\begin{table}[tbp]
%\rowcolors{1}{lightgray}{white}
\begin{tabular}{|c|c|c|}
\hline\hline
The state $B(J^P)$                     & Mass (MeV) & Residue $\lambda~(\mbox{GeV}^3)$  \\
 \hline\hline
$ \Xi_b(\frac{1}{2}^-)(1P)$        & $6190^{+165}_{-145}$ & $ 0.145^{+0.030}_{-0.040}$ \\
$ \Xi_b(\frac{1}{2}^+)(2S)$        & $6190^{+165}_{-145}$ & $ 0.810^{+0.050}_{-0.060}$\\
$\Xi_b^{\prime}(\frac{1}{2}^-)(1P)$& $6195^{+140}_{-150}$ & $ 0.110^{+0.050}_{-0.070} $\\
$\Xi_b^{\prime}(\frac{1}{2}^+)(2S)$& $6195^{+140}_{-150}$ & $ 0.760^{+0.190}_{-0.240} $ \\
$\Xi_b^(\frac{3}{2}^-)(1P)$        & $6200^{+90}_{-105}$  & $ 0.075^{+0.010}_{-0.010}$ \\
$\Xi_b^(\frac{3}{2}^+)(2S)$        & $6200^{+90}_{-105}$  & $ 0.540^{+0.030}_{-0.050}$  \\
\hline\hline
\end{tabular}%
\caption{Results obtained for $ 1P $ and $ 2S $ excitations of the ground state $\Xi_b^-$, $\Xi_b^{\prime}(5935)^-$ and $\Xi_b(5955)^{-}$ baryons.}
\label{tab:results}
\end{table}
The errors in the results are due to the errors of   the input parameters as well as those coming from the variations of the results with respect to the variations of the auxiliary parameters in their working intervals. As an example, in Fig.~\ref{gr:mXib},   we present the dependence of the mass of $\Xi_b(\frac{3}{2}^-)(1P)$ state  on $M^2$ and $s_0$. From this figure, one can see that the requirement of the relatively weak dependence on these parameters is satisfied.  

The masses  of the other possibilities and decay constants for all the states under consideration are determined similarly and presented in Table~\ref{tab:results}. From this table, it follows that the masses of the considered states with $J^P=\frac{1}{2}^{\pm}$ and $J^P=\frac{3}{2}^{\pm}$ are all very close to each other and therefore the mass determination is not enough to identify $\Xi_b(6227)^-$. Hence, in the next section, we extend the analysis and obtain the  widths of  the considered excited states decaying to $\Lambda_b^0 K^-$ and $\Xi_b^0 \pi^-$, which can provide us with the possibility to assign quantum numbers of the $\Xi_b(6227)^-$ state. 

\section{Transitions of the $\Xi_b$ States to $\Lambda_b^0 K^-$ and $\Xi_b^0 \pi^-$}

This section is devoted to the analysis of the transitions $\Xi_b(6227)^-\rightarrow \Lambda_b^0 K^-$ and $\Xi_b(6227)^-\rightarrow\Xi_b^0 \pi^-$ by considering the $\Xi_b(6227)^-$ as $ 1P $ or $ 2S $ excitation state of one of the ground state $\Xi_b^-$, $\Xi_b^{\prime}(5935)^-$, or $\Xi_b(5955)^{-}$ baryon. In calculations, the LCSR method is employed. The starting point of this method is consideration of the correlation function given as
\begin{equation}
\Pi _{(\mu)}(q)=i\int d^{4}xe^{iq\cdot x}\langle K (\pi)(q)|\mathcal{T}\{J_{\Lambda_b^0(\Xi_b^0)
}(x)\bar{J}_{B(\mu) }(0)\}|0\rangle.  \label{eq:CorrF2}
\end{equation}
In this correlation function, $J_{B(\mu)}$ represents one of the currents given in Eqs.~(\ref{Eq:Current1a}) and (\ref{Eq:Current1b}) and the calculation of this correlation function  will be carried on for each current given there, considering again both possibilities of being $1P$ or $2S$ states separately. The index $(\mu)$ is used only for transition of $J=\frac{3}{2}$ state. $\langle K (\pi)(q)|$ is the on-shell $K(\pi)$-meson state with momentum $q$. $J_{\Lambda_b^0}$ and $J_{\Xi_b^0}$ are the interpolating currents of the $J=\frac{1}{2}$ $\Lambda_b^0$ and $\Xi_b^0$ baryons. 

In this part of the study, again the correlation function is firstly calculated in terms of the hadronic parameters. For $J=\frac{1}{2}$ we get
\begin{eqnarray}
\Pi ^{\mathrm{Phys}}(p,q)&=&\frac{\langle 0|J _{\Lambda_b(\Xi_b)}|\Lambda_b(\Xi_b) (p,s)\rangle
}{p^{2}-m_{\Lambda_b(\Xi_b)}^{2}}\langle K(\pi)(q)\Lambda_b(\Xi_b)(p,s)|B
(p^{\prime },s^{\prime })\rangle  \frac{\langle B(p^{\prime },s^{\prime
})|\bar{J}_{B}|0\rangle }{p^{\prime
2}-m^{2}}\nonumber\\
&+&\frac{\langle 0|J _{\Lambda_b(\Xi_b)}|\Lambda_b(\Xi_b) (p,s)\rangle
}{p^{2}-m_{\Lambda_b(\Xi_b)}^{2}}\langle K(\pi)(q)\Lambda_b(\Xi_b)(p,s)|\widetilde{B}
(p^{\prime },s^{\prime })\rangle  \frac{\langle \widetilde{B}(p^{\prime },s^{\prime
})|\bar{J}_{B}|0\rangle }{p^{\prime
2}-{\widetilde{m}}^{2}}
 +\ldots ,  \label{eq:SRDecay1a}
\end{eqnarray} 
\begin{eqnarray}
\Pi ^{\mathrm{Phys}}(p,q)&=&\frac{\langle 0|J _{\Lambda_b(\Xi_b)}|\Lambda_b(\Xi_b) (p,s)\rangle
}{p^{2}-m_{\Lambda_b(\Xi_b)}^{2}}\langle K(\pi)(q)\Lambda_b(\Xi_b)(p,s)|B
(p^{\prime },s^{\prime })\rangle  \frac{\langle B(p^{\prime },s^{\prime
})|\bar{J}_{B}|0\rangle }{p^{\prime
2}-m^{2}}\nonumber\\
&+&\frac{\langle 0|J _{\Lambda_b(\Xi_b)}|\Lambda_b(\Xi_b) (p,s)\rangle
}{p^{2}-m_{\Lambda_b(\Xi_b)}^{2}}\langle K(\pi)(q)\Lambda_b(\Xi_b)(p,s)|B'
(p^{\prime },s^{\prime })\rangle  \frac{\langle B'(p^{\prime },s^{\prime
})|\bar{J}_{B}|0\rangle }{p^{\prime
2}-m'{}^{2}}
 +\ldots ,  \label{eq:SRDecay1b}
\end{eqnarray} 
and for $J=\frac{3}{2}$ we obtain
\begin{eqnarray}
\Pi ^{\mathrm{Phys}}_{\mu}(p,q)&=&\frac{\langle 0|J _{\Lambda_b(\Xi_b)}|\Lambda_b(\Xi_b) (p,s)\rangle
}{p^{2}-m_{\Lambda_b(\Xi_b)}^{2}}\langle K(\pi)(q)\Lambda_b(\Xi_b)(p,s)|B^{*}
(p^{\prime },s^{\prime })\rangle  \frac{\langle B^{*}(p^{\prime },s^{\prime
})|\bar{J}_{B,\mu}|0\rangle }{p^{\prime
2}-m^{*}{}^{2}}\nonumber\\
&+&\frac{\langle 0|J _{\Lambda_b(\Xi_b)}|\Lambda_b(\Xi_b) (p,s)\rangle
}{p^{2}-m_{\Lambda_b(\Xi_b)}^{2}}\langle K(\pi)(q)\Lambda_b(\Xi_b)(p,s)|\widetilde{B}^{*}
(p^{\prime },s^{\prime })\rangle  \frac{\langle \widetilde{B}^{*}(p^{\prime },s^{\prime
})|\bar{J}_{B,\mu}|0\rangle }{p^{\prime
2}-\widetilde{m}^{*}{}^{2}}
 +\ldots ,  \label{eq:SRDecay2a}
\end{eqnarray} 
\begin{eqnarray}
\Pi ^{\mathrm{Phys}}_{\mu}(p,q)&=&\frac{\langle 0|J _{\Lambda_b(\Xi_b)}|\Lambda_b(\Xi_b) (p,s)\rangle
}{p^{2}-m_{\Lambda_b(\Xi_b)}^{2}}\langle K(\pi)(q)\Lambda_b(\Xi_b)(p,s)|B^{*}
(p^{\prime },s^{\prime })\rangle  \frac{\langle B^{*}(p^{\prime },s^{\prime
})|\bar{J}_{B,\mu}|0\rangle }{p^{\prime
2}-m^{*}{}^{2}}\nonumber\\
&+&\frac{\langle 0|J _{\Lambda_b(\Xi_b)}|\Lambda_b(\Xi_b) (p,s)\rangle
}{p^{2}-m_{\Lambda_b(\Xi_b)}^{2}}\langle K(\pi)(q)\Lambda_b(\Xi_b)(p,s)|B^{*}{}'
(p^{\prime },s^{\prime })\rangle  \frac{\langle B^{*}{}'(p^{\prime },s^{\prime
})|\bar{J}_{B,\mu}|0\rangle }{p^{\prime
2}-m^{*}{}'{}^{2}}
 +\ldots .  \label{eq:SRDecay2b}
\end{eqnarray} 
Again note that, in these equations, $B(B^{*})$, $\widetilde{B}(\widetilde{B}^{*})$, and $B'(B^{*}{}')$ represent the ground,  $1P$ and  $2S$ excitated states corresponding to each considered ground state baryon. Here, $p^{\prime}=p+q$ and $p$ are the momenta of these baryons and $\Lambda_b(\Xi_b)$ baryon, respectively. The contributions of higher states and continuum are represented by dots. 

Now, we have some additional matrix elements in Eqs.~(\ref{eq:SRDecay1a})-(\ref{eq:SRDecay2b}). For $J=\frac{1}{2}$ baryons, these matrix elements are determined as 
\begin{eqnarray}
\langle K(\pi)(q)\Lambda_b(\Xi_b)(p,s)|B(p^{\prime },s^{\prime })\rangle &=& g_{B\Lambda_b(\Xi_b)K(\pi)}\bar{u}(p,s) \gamma _5u(p',s'),\nonumber\\
\langle K(\pi)(q)\Lambda_b(\Xi_b)(p,s)|\widetilde{B}(p^{\prime },s^{\prime })\rangle &=& g_{\widetilde{B}\Lambda_b(\Xi_b)K(\pi)}\bar{u}(p,s) u(p',s'),\nonumber\\
\langle K(\pi)(q)\Lambda_b(\Xi_b)(p,s)|B'(p^{\prime },s^{\prime })\rangle &=& g_{B'\Lambda_b(\Xi_b)K(\pi)}\bar{u}(p,s) \gamma_5 u(p',s')\label{eq:Res2},
\end{eqnarray}
and, for $J=\frac{3}{2}$ baryons, they are  parametrized as  
\begin{eqnarray}
\langle K(\pi)(q)\Lambda_b(\Xi_b)(p,s)|B^{*}(p^{\prime },s^{\prime })\rangle &=& g_{B^{*}\Lambda_b(\Xi_b)K(\pi)}\bar{u}(p,s)u_{\mu}(p',s')q^{\mu},\nonumber\\
\langle K(\pi)(q)\Lambda_b(\Xi_b)(p,s)|\widetilde{B}^{*}(p^{\prime },s^{\prime })\rangle &=& g_{\widetilde{B}^{*}\Lambda_b(\Xi_b)K(\pi)}\bar{u}(p,s) \gamma_5 u_{\mu}(p',s')q^{\mu},\nonumber\\
\langle K(\pi)(q)\Lambda_b(\Xi_b)(p,s)|B^{*}{}'(p^{\prime },s^{\prime })\rangle &=& g_{B^{*}{}'\Lambda_b(\Xi_b)K(\pi)}\bar{u}(p,s)u_{\mu}(p',s')q^{\mu}\label{eq:Res3},
\end{eqnarray}
where, in Eqs.~(\ref{eq:Res2}) and (\ref{eq:Res3})  $g$'s with various indices denote the  strong coupling constants of the corresponding baryons with pseudoscalar mesons. 

Inserting these matrix elements into Eqs.~(\ref{eq:SRDecay1a})-(\ref{eq:SRDecay2b})  and applying summations over spins given in Eqs.~(\ref{Dirac}) and (\ref{Rarita}), for physical sides of the correlation function, we get 
\begin{eqnarray}
\Pi ^{\mathrm{Phys}}(p,q)&=&\frac{%
g_{B \Lambda_b(\Xi_b) K(\pi)}\lambda _{\Lambda_b(\Xi_b)}\lambda }{%
(p^{2}-m_{\Lambda_b(\Xi_b)}^{2})(p^{\prime }{}^{2}-m^{ 2})}(\slashed p+m_{\Lambda_b(\Xi_b)})\gamma _{5}\left( \slashed p'+m \right) -\frac{g_{\widetilde{B}\Lambda_b(\Xi_b) K(\pi)}\lambda _{\Lambda_b(\Xi_b)}%
\widetilde{\lambda} }{(p^{2}-m_{\Lambda_b(\Xi_b)}^{2})(p^{\prime }{}^{2}-\widetilde{m}^{2})}(\slashed p+m_{\Lambda_b(\Xi_b)})  \notag \\
&\times &\left( \slashed p'+\widetilde{m}\right) \gamma _{5}+\ldots , \label{eq:SRDecayPhys1a}
\end{eqnarray}%
\begin{eqnarray}
\Pi ^{\mathrm{Phys}}(p,q)&=&\frac{%
g_{B \Lambda_b(\Xi_b) K(\pi)}\lambda _{\Lambda_b(\Xi_b)}\lambda }{%
(p^{2}-m_{\Lambda_b(\Xi_b)}^{2})(p^{\prime }{}^{2}-m^{ 2})}(\slashed p+m_{\Lambda_b(\Xi_b)})\gamma _{5}\left( \slashed p'+m \right) +\frac{%
g_{B' \Lambda_b(\Xi_b) K(\pi)}\lambda _{\Lambda_b(\Xi_b)}\lambda'}{%
(p^{2}-m_{\Lambda_b(\Xi_b)}^{2})(p^{\prime }{}^{2}-m'{}^{ 2})}(\slashed p+m_{\Lambda_b(\Xi_b)})\notag \\
&\times &\gamma _{5}\left( \slashed p'+m' \right)+\ldots ,\label{eq:SRDecayPhys1b}
\end{eqnarray}%
\begin{eqnarray}
\Pi _{\mu }^{\mathrm{Phys}}(p,q)&=&-\frac{g_{B^{*} \Lambda_b(\Xi_b) K(\pi)}\lambda
_{\Lambda_b(\Xi_b)}\lambda}{(p^{2}-m_{\Lambda_b(\Xi_b)}^{2})(p^{\prime
2}-m^{*}{}^{2})}q^{\alpha }(\slashed p+m_{\Lambda_b(\Xi_b)})\left( \slashed p'+m^{*}\right) T_{\alpha \mu }+\frac{g_{\widetilde{B}^{*} \Lambda_b(\Xi_b) K(\pi)}\lambda
_{\Lambda_b(\Xi_b)}\widetilde{\lambda}^{*}}{(p^{2}-m_{\Lambda_b(\Xi_b)}^{2})(p^{\prime }{}^{2}-%
\widetilde{m}^{*}{}^{2})} \notag \\
&\times & q^{\alpha }(\slashed p+m_{\Lambda_b(\Xi_b)})  \gamma _{5}\left( \slashed p'+\widetilde{m}^{*}\right) T_{\alpha \mu }\gamma _{5}+\ldots,\label{eq:SRDecayPhys2a}
\end{eqnarray}
\begin{eqnarray}
\Pi _{\mu }^{\mathrm{Phys}}(p,q)&=&-\frac{g_{B^{*} \Lambda_b(\Xi_b) K(\pi)}\lambda
_{\Lambda_b(\Xi_b)}\lambda}{(p^{2}-m_{\Lambda_b(\Xi_b)}^{2})(p^{\prime
2}-m^{*}{}^{2})}q^{\alpha }(\slashed p+m_{\Lambda_b(\Xi_b)})\left( \slashed p'+m^{*}\right) T_{\alpha \mu }-\frac{g_{B^{*}{}' \Lambda_b(\Xi_b) K(\pi)}\lambda
_{\Lambda_b(\Xi_b)}\lambda^{*}{}'}{(p^{2}-m_{\Lambda_b(\Xi_b)}^{2})(p^{\prime
2}-m^{*}{}'^{2})}\notag\\
&\times & q^{\alpha }(\slashed p+m_{\Lambda_b(\Xi_b)})\left( \slashed p'+m^{*}{}'\right) T_{\alpha \mu }+\ldots\label{eq:SRDecayPhys2b}.
\end{eqnarray}%
 The function $T_{\alpha\mu}$ is given as
\begin{eqnarray}
&&T_{\alpha \mu }(p)=g_{\alpha \mu }-\frac{1}{3}\gamma _{\alpha }\gamma
_{\mu }-\frac{2}{3m^{2}}p_{\alpha }p_{\mu }+\frac{1}{3m}\left[ p_{\alpha }\gamma _{\mu }-p_{\mu
}\gamma _{\alpha }\right] .
\end{eqnarray}%   
To suppress the contributions coming from the higher states and continuum, a double Borel transformation with respect to $-p^2$ and $-p'^2$ is performed. As a  result, we get  
\begin{eqnarray}
{\cal B} \Pi ^{\mathrm{Phys}}(p,q)&=&g_{B \Lambda_b(\Xi_b) K(\pi)}\lambda _{\Lambda_b(\Xi_b)}\lambda
   e^{-m^{2}/M_{1}^{2}}e^{-m_{\Lambda_b(\Xi_b)}^{2}/M_{2}^{2}}\left\{ %
\slashed q\slashed p\gamma _{5}-m_{\Lambda_b(\Xi_b)}\slashed q\gamma _{5}+\left(
m-m_{\Lambda_b(\Xi_b)}\right) \slashed p\gamma _{5}\right.  \notag \\
&+&\left. \left[ m_{K(\pi)}^{2}-m(m-m_{\Lambda_b(\Xi_b)})\right]
\gamma _{5}\right\} +g_{\widetilde{B} \Lambda_b(\Xi_b) K(\pi)}\lambda _{\Lambda_b(\Xi_b)}\widetilde{\lambda}e^{-\widetilde{m}^{2}/M_{1}^{2}}e^{-m_{\Lambda_b(\Xi_b)}^{2}/M_{2}^{2}}  \notag \\
&&\times \left\{ \slashed q\slashed p\gamma _{5}-m_{\Lambda_b(\Xi_b)}\slashed %
q\gamma _{5}-\left( \widetilde{m}+m_{\Lambda_b(\Xi_b)}\right) \slashed p\gamma
_{5}+\left[ m_{K(\pi)}^{2}-\widetilde{m}(\widetilde{m}+m_{\Lambda_b(\Xi_b)})\right]
\gamma _{5}\right\} ,  \label{eq:CFunc1/2a}
\end{eqnarray}
\begin{eqnarray}
{\cal B} \Pi ^{\mathrm{Phys}}(p,q)&=&g_{B \Lambda_b(\Xi_b) K(\pi)}\lambda _{\Lambda_b(\Xi_b)}\lambda
 e^{-m^{2}/M_{1}^{2}}e^{-m_{\Lambda_b(\Xi_b)}^{2}/M_{2}^{2}}\left\{ %
\slashed q\slashed p\gamma _{5}-m_{\Lambda_b(\Xi_b)}\slashed q\gamma _{5}+\left(
m-m_{\Lambda_b(\Xi_b)}\right) \slashed p\gamma _{5}\right.  \notag \\
&+&\left. \left[ m_{K(\pi)}^{2}-m(m-m_{\Lambda_b(\Xi_b)})\right]
\gamma _{5}\right\} +g_{B' \Lambda_b(\Xi_b) K(\pi)}\lambda _{\Lambda_b(\Xi_b)}\lambda'
   e^{-m'{}^{2}/M_{1}^{2}}e^{-m_{\Lambda_b(\Xi_b)}^{2}/M_{2}^{2}} \notag \\
&&\times \left\{ \slashed q\slashed p\gamma _{5}-m_{\Lambda_b(\Xi_b)}\slashed q\gamma _{5}+\left(
m'-m_{\Lambda_b(\Xi_b)}\right) \slashed p\gamma _{5}+\left[ m_{K(\pi)}^{2}-m'(m'-m_{\Lambda_b(\Xi_b)})\right]
\gamma _{5}\right\} ,  \label{eq:CFunc1/2b}
\end{eqnarray}%
\begin{eqnarray}
{\cal B} \Pi _{\mu }^{\mathrm{Phys}}(p,q)&=&-g_{B^{*} \Lambda_b(\Xi_b) K(\pi)}\lambda _{\Lambda_b(\Xi_b)}\lambda^{* }e^{-m^{*}{}%
^{2}/M_{1}^{2}}e^{-m_{\Lambda_b(\Xi_b)}^{2}/M_{2}^{2}}q^{\alpha }  (\slashed p+m_{\Lambda_b(\Xi_b)})\left( \slashed p'+%
m^{*}\right) T_{\alpha \mu }\notag \\
&+&g_{\widetilde{B}^{*} \Lambda_b(\Xi_b) K(\pi)}\lambda _{\Lambda_b(\Xi_b)}\widetilde{\lambda}^{*}  
 e^{-\widetilde{m}^{*}{}^2/M_{1}^{2}}e^{-m_{\Lambda_b(\Xi_b)}^{2}/M_{2}^{2}}q^{\alpha
}(\slashed p+m_{\Lambda_b(\Xi_b)})\gamma _{5}\left( \slashed p'+\widetilde{m}^{*} \right)
T_{\alpha \mu }\gamma _{5}. \label{eq:CFunc3/2a}
\end{eqnarray}
\begin{eqnarray}
{\cal B} \Pi _{\mu }^{\mathrm{Phys}}(p,q)&=&-g_{B^{*} \Lambda_b(\Xi_b) K(\pi)}\lambda _{\Lambda_b(\Xi_b)}\lambda^{* }e^{-m^{*}{}%
^{2}/M_{1}^{2}}e^{-m_{\Lambda_b(\Xi_b)}^{2}/M_{2}^{2}}q^{\alpha }  (\slashed p+m_{\Lambda_b(\Xi_b)})\left( \slashed p'+%
m^{*}\right) T_{\alpha \mu }\notag \\
&-&g_{B^{*}{}' \Lambda_b(\Xi_b) K(\pi)}\lambda _{\Lambda_b(\Xi_b)}\lambda^{*}{}' e^{-m^{*}{}'{}%
^{2}/M_{1}^{2}}e^{-m_{\Lambda_b(\Xi_b)}^{2}/M_{2}^{2}}q^{\alpha }  (\slashed p+m_{\Lambda_b(\Xi_b)})\left( \slashed p'+%
m^{*}{}'\right) T_{\alpha \mu },  \label{eq:CFunc3/2b}
\end{eqnarray}
where, $M_1^2$ and $M_2^2$ are Borel parameters in initial and final channels, respectively.  In these equations  
${\cal B}\Pi_{(\mu)}^{\mathrm{Phys}}(p,q)$ stands for the Borel transformed form of the  $\Pi_{(\mu) }^{\mathrm{Phys}}(p,q)$ function. These results contain different Lorentz structures from which we can get the sum rules to obtain the strong coupling constants under question. In the $J=\frac{1}{2}$ scenario, we use $\slashed q \slashed p\gamma_5$ and $\slashed p \gamma_5$ structures to obtain the coupling constants for each possibility that the state $\Xi_b(6227)$ may become. For obtaining the relevant coupling constants for  $J=\frac{3}{2}$ baryon, the Lorentz structures  $\slashed q \slashed p\gamma_{\mu}$ and $\slashed q q_{\mu}$ are considered. 

For both $J^P=\frac{1}{2}^{\pm}$ and  $J^P=\frac{3}{2}^{\pm}$ scenarios, we also need to calculate the theoretical sides of the correlation function, Eq.~(\ref{eq:CorrF2}), with usage of the related interpolating currents, explicitly. Similar to the previous case after the possible contractions made using Wick's theorem between the quark fields of the interpolating currents, the results are expressed in terms of light and heavy quark propagators. Besides the propagators, we need matrix elements of remaining quark field operators  between the  $K(\pi)$ meson and the vacuum. These matrix elements, whose common forms can be written as $\langle K(\pi)(q)|\bar{q}(x)\Gamma q(y)|0\rangle$ or $\langle K(\pi)(q)|\bar{q}(x)\Gamma G_{\mu\nu} q(y)|0\rangle$ with $\Gamma$ and $G_{\mu\nu}$ being full set of Dirac matrices and gluon field strength tensor respectively, are parameterized in terms of $K(\pi)$-meson distribution amplitudes (DAs). Nonperturbative contributions are attained  by exploiting these matrix elements as inputs in the calculations. The explicit form of these matrix elements are present in Refs.~\cite{Ball:2006wn,Belyaev:1994zk,Ball:2004ye,Ball:2004hn}. 

Again, considering the same structures in the physical  and the theoretical sides and matching the coefficients of the same structures in both sides, performing the  Borel transformation and continuum subtraction using quark hadron duality assumption, we obtain the QCD sum rules for the relevant coupling constants as
\begin{eqnarray}
{\cal B}\Pi_1^{\mathrm{OPE}}&=&g_{{B}\Lambda_b(\Xi_b) K(\pi)} \lambda_{\Lambda_b(\Xi_{b}) } \lambda e^{-\frac{m^2}{M_1^2}}e^{-\frac{m_{\Lambda_b(\Xi_b)}^2}{M_2^2}}+g_{\widetilde{B}\Lambda_b(\Xi_b) K(\pi)}\lambda_{\Lambda_b(\Xi_{b}) } \widetilde{\lambda} e^{-\frac{\widetilde{m}^{2}}{M_{1}^{2}}}e^{-\frac{m_{\Lambda_b(\Xi_b)}^2}{M_2^2}} \nonumber\\
{\cal B}\Pi_2^{\mathrm{OPE}}&=&g_{{B}\Lambda_b(\Xi_b) K(\pi)} \lambda_{\Lambda_b(\Xi_{b}) } \lambda  e^{-\frac{m^2}{M_1^2}} e^{-\frac{m_{\Lambda_b(\Xi_b)}^2}{M_2^2}}(m-m_{\Lambda_b(\Xi_{b})})-g_{\widetilde{B}\Lambda_b(\Xi_b) K(\pi)}\lambda_{\Lambda_b(\Xi_{b}) } \widetilde{\lambda} e^{-\frac{\widetilde{m}^{2}}{M_{1}^{2}}}e^{-\frac{m_{\Lambda_b(\Xi_b)}^2}{M_2^2}}\nonumber\\
&\times&(\widetilde{m}+m_{\Lambda_b(\Xi_{b})})\label{eq:couplingpair1/2a},
\end{eqnarray}
\begin{eqnarray}
{\cal B}\Pi_1^{\mathrm{OPE}}&=&g_{{B}\Lambda_b(\Xi_b) K(\pi)} \lambda_{\Lambda_b(\Xi_{b}) } \lambda  e^{-\frac{m^2}{M_1^2}}e^{-\frac{m_{\Lambda_b(\Xi_b)}^2}{M_2^2}}+g_{B'\Lambda_b(\Xi_b) K(\pi)}\lambda_{\Lambda_b(\Xi_{b}) } \lambda' e^{-\frac{m'^{2}}{M_{1}^{2}}}e^{-\frac{m_{\Lambda_b(\Xi_b)}^2}{M_2^2}} \nonumber\\
{\cal B}\Pi_2^{\mathrm{OPE}}&=&g_{B\Lambda_b(\Xi_b) K(\pi)}  \lambda_{\Lambda_b(\Xi_{b}) } \lambda e^{-\frac{m^2}{M_1^2}}e^{-\frac{m_{\Lambda_b(\Xi_b)}^2}{M_2^2}}(m-m_{\Lambda_b(\Xi_{b})})+g_{B'\Lambda_b(\Xi_b) K(\pi)}\lambda_{\Lambda_b(\Xi_{b}) } \lambda' e^{-\frac{m'{}^{2}}{M_{1}^{2}}}e^{-\frac{m_{\Lambda_b(\Xi_b)}^2}{M_2^2}}\nonumber\\
&\times&(m'-m_{\Lambda_b(\Xi_{b})})\label{eq:couplingpair1/2b},
\end{eqnarray}
\begin{eqnarray}
{\cal B}\Pi_{1}^{*}{}^{\mathrm{OPE}}&=&-g_{B^{*}\Lambda_b(\Xi_b) K(\pi)}  \lambda_{\Lambda_b(\Xi_{b}) } \lambda^{*} \frac{[(m^{*}+m_{\Lambda_b(\Xi_{b})})^2-m_{K(\pi)}^2]}{6m^{*}}e^{-\frac{m^{*}{}^2}{M_1^2}}e^{-\frac{m_{\Lambda_b(\Xi_b)}^2}{M_2^2}}+g_{\widetilde{B}^{*}\Lambda_b(\Xi_b) K(\pi)}\lambda_{\Lambda_b(\Xi_{b}) } \widetilde{\lambda}^{*} \nonumber\\&\times& \frac{[(\widetilde{m}^{*}-m_{\Lambda_b(\Xi_{b})})^2-m_{K(\pi)}^2]}{6\widetilde{m}^{*}}e^{-\frac{\widetilde{m}^{*}{}^{2}}{M_{1}^{2}}}e^{-\frac{m_{\Lambda_b(\Xi_b)}^2}{M_2^2}}\nonumber\\
{\cal B}\Pi_{2}^{*}{}^{\mathrm{OPE}}&=&-g_{B^{*}\Lambda_b(\Xi_b) K(\pi)}  \lambda_{\Lambda_b(\Xi_{b}) } \lambda^{*} \frac{[m^{*}{}^2+m_{\Lambda_b(\Xi_{b})}^2-m^{*} m_{\Lambda_b(\Xi_{b})}-m_{K(\pi)}^2]m_{\Lambda_b(\Xi_{b})}}{3m^{*}{}^2}e^{-\frac{m^{*}{}^2}{M_1^2}}e^{-\frac{m^2_{\Lambda_b(\Xi_b)}}{M_2^2}}\nonumber\\&-&g_{\widetilde{B}^{*}\Lambda_b(\Xi_b) K(\pi)}\lambda_{\Lambda_b(\Xi_{b}) } \widetilde{\lambda}^{*}
\frac{[\widetilde{m}^{*}{}^2+m_{\Lambda_b(\Xi_{b})}^2+\widetilde{m}^{*}m_{\Lambda_b(\Xi_{b})}-m_{K(\pi)}^2]m_{\Lambda_b(\Xi_{b})}}{3\widetilde{m}^{*}{}^2} e^{-\frac{\widetilde{m}^{*}{}^{2}}{M_{1}^{2}}}e^{-\frac{m_{\Lambda_b(\Xi_b)}^2}{M_2^2}}\label{eq:couplingpair3/2a},
\end{eqnarray}
\begin{eqnarray}
{\cal B}\Pi_{1}^{*}{}^{\mathrm{OPE}}&=&-g_{B^{*}\Lambda_b(\Xi_b) K(\pi)}  \lambda_{\Lambda_b(\Xi_{b}) }\lambda^{*} \frac{[(m^{*}+m_{\Lambda_b(\Xi_{b})})^2-m_{K(\pi)}^2]}{6m^{*}} e^{-\frac{m^{*}{}^2}{M_1^2}}e^{-\frac{m_{\Lambda_b(\Xi_b)}^2}{M_2^2}}-g_{B^{*}{}'\Lambda_b(\Xi_b) K(\pi)} \lambda_{\Lambda_b(\Xi_{b}) }\lambda^{*}{}'\nonumber\\&\times& \frac{[(m^{*}{}'+m_{\Lambda_b(\Xi_{b})})^2-m_{K(\pi)}^2]}{6m^{*}{}'} e^{-\frac{m^{*}{}'{}^{2}}{M_{1}^{2}}}e^{-\frac{m_{\Lambda_b(\Xi_b)}^2}{M_2^2}}\nonumber\\
{\cal B}\Pi_{2}^{*}{}^{\mathrm{OPE}}&=&-g_{B^{*}\Lambda_b(\Xi_b) K(\pi)}  \lambda_{\Lambda_b(\Xi_{b}) } \lambda^{*} \frac{[m^{*}{}^2+m_{\Lambda_b(\Xi_{b})}^2-m^{*} m_{\Lambda_b(\Xi_{b})}-m_{K(\pi)}^2]m_{\Lambda_b(\Xi_{b})}}{3m^{*}{}^2}e^{-\frac{m^{*}{}^2}{M_1^2}}e^{-\frac{m_{\Lambda_b(\Xi_b)}^2}{M_2^2}}\nonumber\\&-&g_{B^{*}{}'\Lambda_b(\Xi_b) K(\pi)}\lambda_{\Lambda_b(\Xi_{b}) } \lambda^{*}{}'
\frac{[m^{*}{}'{}^2+m_{\Lambda_b(\Xi_{b})}^2-m^{*}{}' m_{\Lambda_b(\Xi_{b})}-m_{K(\pi)}^2]m_{\Lambda_b(\Xi_{b})}}{3m^{*}{}'{}^2} e^{-\frac{m^{*}{}'{}^{2}}{M_{1}^{2}}}e^{-\frac{m_{\Lambda_b(\Xi_b)}^2}{M_2^2}}.\label{eq:couplingpair3/2b}
\end{eqnarray}
where ${\cal B}\Pi_{1}^{\mathrm{OPE}}({\cal B}\Pi_{1}^{*}{}^{\mathrm{OPE}})$ and $ {\cal B}\Pi_{2}^{\mathrm{OPE}}({\cal B}\Pi_{2}^{*}{}^{\mathrm{OPE}})$ represent the Borel transformed coefficients of the $\slashed q \slashed p\gamma_5(\slashed q \slashed p\gamma_{\mu})$ and $\slashed p \gamma_5(\slashed q q_{\mu})$ structures of theoretical sides for $J=\frac{1}{2}(\frac{3}{2})$ case. As examples, we present the explicit expressions of the QCD sides obtained for the decay of $ J=\frac{1}{2} $, $\Xi_b^-$ baryon to $\Xi_b^0$ and $\pi^-$ states in the Appendix.

\begin{figure}[h!]
\begin{center}
\includegraphics[totalheight=5cm,width=7cm]{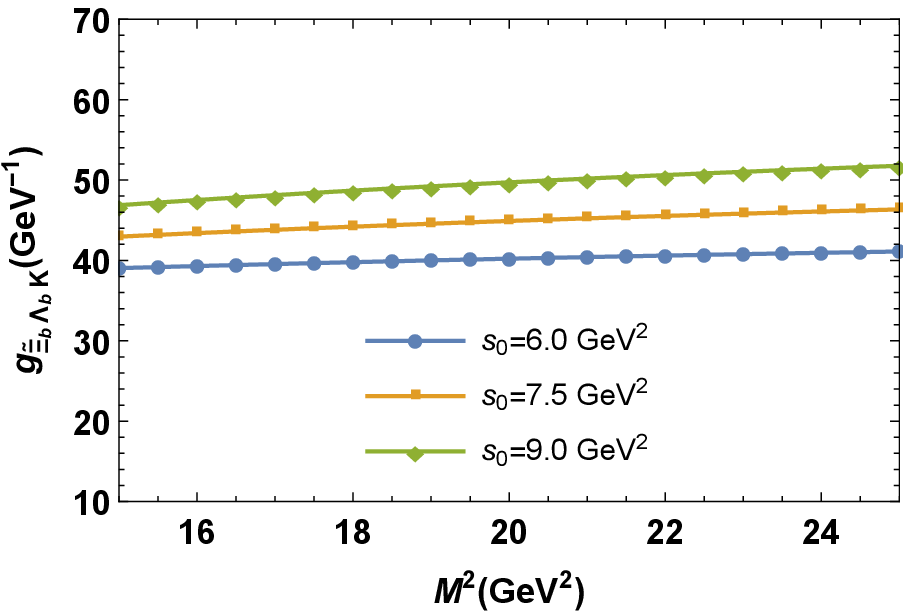}
\includegraphics[totalheight=5cm,width=7cm]{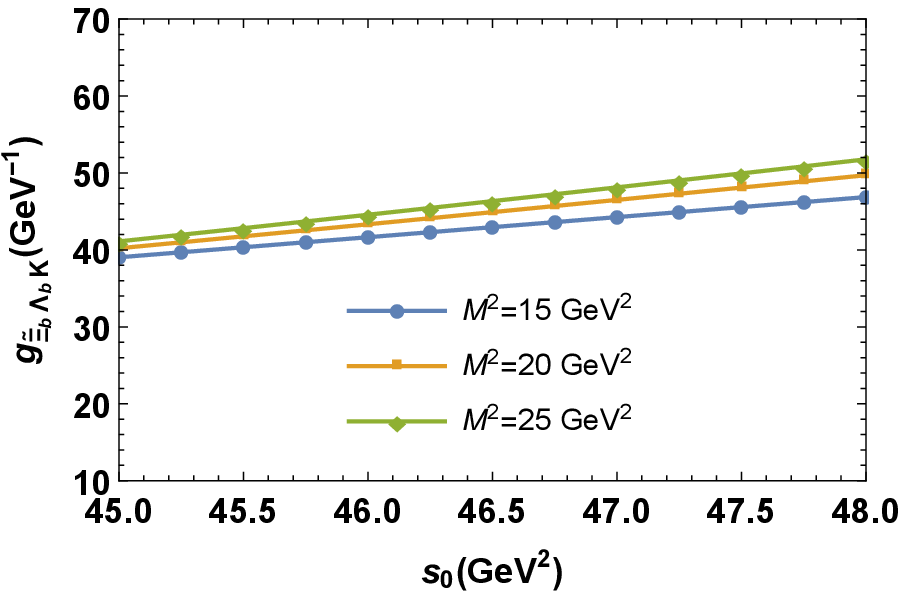}
\end{center}
\caption{\textbf{Left:} The coupling constant $g_{\widetilde{\Xi}_b\Lambda_bK}$ of the $ 1P $ excitation of $\Xi_b(5955)^{-}$ baryon to $\Lambda_b K$ vs Borel parameter $M^2$.
\textbf{Right:} The coupling constant $g_{\widetilde{\Xi}_b\Lambda_bK}$ of the $ 1P $ excitation of $\Xi_b(5955)^{-}$ baryon to $\Lambda_b K$ vs threshold parameter $s_0$. }
\label{gr:gXib}
\end{figure}

 To perform the calculations for the coupling constants, the numerical values of the input parameters presented in Table~\ref{tab:Param} are used. Since the masses of the considered baryons are close to each other, we choose $ M_1^2=M_2^2 $ in   
\begin{eqnarray}
M^2=\frac{M_1^2 M_2^2}{M_1^2+M_2^2},
\end{eqnarray}
entering the calculations, which leads to
\begin{eqnarray}
M_1^2=M_2^2=2M^2.
\end{eqnarray}
 Moreover, for all of the auxiliary parameters, we adopt the values obtained in the mass and decay constant calculations only with one exception. Using the OPE series convergence and pole dominance conditions for working region of $M^2$, in this part, we obtain 
\begin{eqnarray}
15~\mbox{GeV}^2 \leq M^2\leq 25~\mbox{GeV}^2.
\end{eqnarray}
Again to illustrate the sensitivity of the results to the auxiliary parameters, we pick out the coupling constant $g_{\widetilde{\Xi}_b\Lambda_bK}$ for the transition of $ 1P $ excitation of $\Xi_b(5955)^{-}$ baryon to $\Lambda_b$ and $K$ final states and present the dependence of the corresponding coupling constant on   $M^2$ and $s_0$ in  Fig.~\ref{gr:gXib}.

Similarly, we perform the analyses for all the coupling constants under consideration. Our final results for the relevant coupling constants are presented in Table~\ref{tab:decayresults}. The obtained coupling constants are used to extract the corresponding decay widths. The decay width formulas for $ 1P $ and $ 2S $ excitations for the $J=\frac{1}{2}$ cases are
\begin{eqnarray}
\Gamma \left( \widetilde{B}\rightarrow \Lambda_b(\Xi _{b})K(\pi)\right) =\frac{%
g_{\widetilde{B}\Lambda_b(\Xi_b) K(\pi)}^{2}}{8\pi \widetilde{m}^{2}}\left[ (\widetilde{m}%
+m_{\Lambda_b(\Xi _{b})})^{2}-m_{K(\pi)}^{2}\right]  f(\widetilde{m},m_{\Lambda_b(\Xi _{b})},m_{K(\pi)}),
\end{eqnarray}%
and
\begin{eqnarray}
\Gamma \left( B'\rightarrow \Lambda_b(\Xi _{b})K(\pi)\right)  &=&%
\frac{g_{B'\Lambda_b(\Xi_b) K(\pi)}^{2}}{8\pi m'{}^{ 2}}\left[ (m'-m_{\Lambda_b(\Xi _{b})})^{2}-m_{K(\pi)}^{2}\right]  f(m' ,m_{\Lambda_b(\Xi _{b})},m_{K(\pi)}).
\end{eqnarray}

The similar decay width expressions for the $J=\frac{3}{2}$ case are 
\begin{eqnarray}
\Gamma (\widetilde{B}^{*} &\rightarrow &\Lambda_b(\Xi_b) K(\pi))=\frac{%
g_{\widetilde{B}^{*}\Lambda_b(\Xi_b) K(\pi)}^{2}}{24\pi \widetilde{m}^{*}{}^{2}}\left[(\widetilde{m}^{*}%
-m_{\Lambda_b(\Xi_b) })^{2}-m_{K(\pi)}^{2}\right]  f^3(\widetilde{m}^{*},m_{\Lambda_b(\Xi_b)},m_{K(\pi)}),
\end{eqnarray}%
and
\begin{eqnarray}
\Gamma (B^{ * }{}' &\rightarrow &\Lambda_b(\Xi_b) K(\pi))=\frac{%
g_{B^{*}{}'\Lambda_b(\Xi_b) K(\pi)}^{2}}{24\pi m^{*}{}'{}^2}\left[ (m^{*}{}'
+m_{\Lambda_b(\Xi_b)})^{2}-m_{K(\pi)}^{2}\right]  f^3(m^{*}{}',m_{\Lambda_b(\Xi_b)},m_{K(\pi)}).
\end{eqnarray}
The function $f(x,y,z)$ appearing in the decay width equations is defined as
\begin{equation*}
f(x,y,z)=\frac{1}{2x}\sqrt{%
x^{4}+y^{4}+z^{4}-2x^{2}y^{2}-2x^{2}z^{2}-2y^{2}z^{2}}.
\end{equation*}

The numerical results  for the decay widths are also  presented in Table~\ref{tab:decayresults}. In this table, we also present the total width values as a sum of the considered transitions in each case. Comparing our results given in Table~\ref{tab:decayresults} to that of the experimental width, which is $\Gamma_{\Xi_b(6227)^-}=18.1\pm5.4\pm 1.8$~MeV~\cite{Aaij:2018yqz}, it can be seen that the state with $J^P=\frac{3}{2}^-$ scenario nicely reproduces the experimental width value.    
\begin{table}[]
\begin{tabular}{|c|c|c|c|c|c|c|}
\hline
 The state $B(J^P)$ & $g_{B\Lambda_b^0 K^-}$ &$g_{B \Xi_b^0 \pi^-}$  & $\Gamma(B \rightarrow \Lambda_b^0 K^-)~(\mbox{MeV})$&$\Gamma(B\rightarrow \Xi_b^0 \pi^-)~(\mbox{MeV})$ & Total Width $\Gamma~(\mbox{MeV})$  \\ \hline\hline
 $\Xi_b(\frac{1}{2}^-)~(1P)$ &$1.8\pm 0.4$ &$0.010\pm 0.002$ &$130\pm 30$ &$0.006\pm0.002$ & $130\pm30$\\ 
$\Xi_b(\frac{1}{2}^+)~(2S)$ &$6.7\pm 1.5$ &$0.06\pm0.01$  &$1.1\pm 0.3$   &$0.00016\pm0.00005$&$1.1\pm 0.3$\\ \hline
$\Xi'_b(\frac{1}{2}^-)~(1P)$&$0.5\pm0.1$ &$0.8\pm0.2$   &$12\pm 4$   &$32\pm 9$ &$44\pm10$\\ 
$\Xi'_b(\frac{1}{2}^+)~(2S)$&$1.6\pm0.4$ &$5.2\pm1.1$   &$0.07\pm0.02$&$1.4\pm0.4$  &$1.5\pm0.4$\\ \hline
$\Xi_b(\frac{3}{2}^-)~(1P)$ &$44\pm 10~\mbox{GeV}^{-1}$&$77\pm6~\mbox{GeV}^{-1}$ &$1.6\pm0.5$ &$15\pm3$ &$17\pm3$\\ 
$\Xi_b(\frac{3}{2}^+)~(2S)$ &$19\pm4~\mbox{GeV}^{-1}$ &$35\pm4~\mbox{GeV}^{-1}$ &$0.4\pm0.1$ &$3.2\pm0.8$ & $3.6\pm0.8$\\ \hline
\end{tabular}
\caption{Coupling constant and decay width results obtained for $ 1P $ and $ 2S $ excitations of the ground state $\Xi_b^-$, $\Xi_b^{\prime}(5935)^-$ and $\Xi_b(5955)^{-}$ baryons.}
\label{tab:decayresults}
\end{table}

At the end of this section, we would like to have a comment on the  heavy quark symmetry partners. In heavy quark effective theory  (HQET), the heavy quark decouple with light quarks in the leading inverse heavy quark mass expansion. Therefore, properties of baryons with one heavy quark are determined with the properties of light quarks (so called diquark). The properties of $P$-wave baryons and their interpolating currents were systematically studied in \cite{Chen:2007xf,Chen:2015kpa}. Using the same notations given in these studies, the heavy baryons belonging to the baryon multiplet are characterized by the set of the quantum numbers $[F,j_l,s_l,\rho/\lambda] $, where $F$ means antitriplet or sextet representation of $ SU(3)_F, j_l $ and $ s_l $ are the total angular and spin momenta of the diquark,  and $\rho$ and $\lambda$ denote the  $\rho$ type $(l_\rho=1,l_\lambda=0) $ and $\lambda $ type  $( l_\rho=0,l_\lambda=1)$, respectively. Here, $ l_\rho$ is the orbital angular momentum between two light quarks and $l_\lambda$ is the angular momentum between heavy quark and diquark. As a result, for the  antitriplet negative parity baryons for instance, the following heavy quark symmetry partners are obtained:
\begin{eqnarray}
\Lambda_b(\frac{1}{2}^-), ~\Xi_b(\frac{1}{2}^-) ~~~~~~[\bar{3}_F,0,1,\rho], \nonumber\\
\Lambda_b(\frac{1}{2}^-,\frac{3}{2}^-), ~\Xi_b(\frac{1}{2}^-,\frac{3}{2}^-) ~~~~~~[\bar{3}_F,1,1,\rho], \nonumber\\
\Lambda_b(\frac{3}{2}^-,\frac{5}{2}^-), ~\Xi_b(\frac{3}{2}^-,\frac{5}{2}^-) ~~~~~~[\bar{3}_F,2,1,\rho], \nonumber\\
\Lambda_b(\frac{1}{2}^-,\frac{3}{2}^-), ~\Xi_b(\frac{1}{2}^-,\frac{3}{2}^-) ~~~~~~[\bar{3}_F,1,0,\lambda].
\end{eqnarray}
The strong decay widths of these states are calculated in \cite{Chen:2007xf}. 

\section{Conclusion}
The present work covers an investigation on the nature of the recently observed $\Xi_b(6227)^-$ baryon. To get the information about the possible quantum numbers of this state, three different ground state particles, namely $\Xi_b^-$, $\Xi_b^{'}(5935)$, and $\Xi_b(5955)^{-}$ are considered. The investigations were conducted focusing on their angular-orbital  $ 1P $ and $ 2S $ excitations. With this orientation, we first calculated the corresponding masses  and decay constants for all these excited states using the two-point QCD sum rule formalism. It is obtained that the sum rules predictions on the masses for all the considered scenarios are very consistent with the experimental mass value. Hence, only mass considerations are not enough to determine the quantum numbers of this state.   Therefore, to get extra information about possible quantum numbers, we also studied the strong decays of all the considered states to $\Lambda_b^0 K^-$ and $\Xi_b^-\pi$. To this end, we applied the  LCSR and obtained the coupling constants for all the possible transitions. These coupling constants were then used to get the corresponding decay widths. From the results on decay widths, it is obtained that the angular-orbital excited $1P$ state with $J^P=\frac{3}{2}^-$ has reproduced well the experimental value of the width. This result indicates that  the state $\Xi_b(6227)^-$ has the quantum numbers  $J^P=\frac{3}{2}^-$.

%%%%%%%%%%%%%%%%%%%%%%%%%%%%%%%%%%%%%%%%%%%%%%%%%%%%%%%%%%%%

\section*{ACKNOWLEDGMENTS}

H. S. thanks Kocaeli University for the partial financial support through the grant BAP 2018/070.

\appendix*
\label{sec:App}
\section{QCD Sides of the Correlation Functions}

In this appendix, as examples, we present the results of the QCD sides  for  the $\Xi_b(6227)^-\rightarrow\Xi_b^0 \pi^-$  transition, i.e. the functions ${\cal B}\Pi_1^{\mathrm{OPE}}$ and ${\cal B}\Pi_2^{\mathrm{OPE}}$. They are obtained as

\begin{eqnarray}\label{magneticmoment2}
{\cal B}\Pi_{1(2)}^{\mathrm{OPE}}&=&
\int_{m_{b}^{2}}^{s_{0}}e^{-{\frac{s}{M^{2}}}-\frac{m_{\pi}^{2}}{4M^2}}\rho_{1(2)}(s)ds+e^{{-\frac{m_b^2}{M^{2}}}-\frac{m_{\pi}^{2}}{4M^2}}\Gamma_{1(2)},
\end{eqnarray}
where the expressions, $\rho_{1(2)}(s)$ and $\Gamma_{1(2)}$ are given as:

\begin{eqnarray}
\rho_1(s)&=& \frac{1}{384 m_b^2 \pi^2} (\beta -1)\Bigg[ f_{\pi} \Big(2 (5 + \beta) m_s \big[m_b^2 (-2 \psi_{10} + \psi_{00} (-4 + M^2)) + 2 \psi_{00} s] (\zeta_1 - 2\zeta_2) \ln(\frac{\Lambda^2}{m_b^2}) + m_b^2 \Big(-4 \psi_{10} m_b \zeta_4 \nonumber\\
&+& 4 \beta \psi_{10} m_b \zeta_4 - 4 \psi_{21} m_b \zeta_4 + 
 4 \beta \psi_{21} m_b \zeta_4 + 20 \psi_{10} m_s \zeta_4 + 4 \beta \psi_{10} m_s \zeta\zeta_4 + 20 \psi_{21} m_s \zeta_4 + 4 \beta \psi_{21} m_s \zeta_4 + 20 \psi_{10} m_b \zeta_1 \nonumber\\
 &+& 28 \beta \psi_{10} m_b \zeta_1 + 20 \psi_{21} m_b \zeta_1 + 28 \beta \psi_{21} m_b \zeta_1 + 30 \psi_{00} m_s \zeta_1 + 6 \beta \psi_{00} m_s \zeta_1 + 30 \gamma_E \psi_{00} m_s \zeta_1 + 6 \beta \gamma_E \psi_{00} m_s \zeta_1 \nonumber\\
 &+& 10 \psi_{01} m_s \zeta_1 +2 \beta \psi_{01} m_s \zeta_1 - 25 \psi_{02} m_s \zeta_1 - 5 \beta \psi_{02} m_s \zeta_1 -10 \gamma_E \psi_{02} m_s \zeta_1 - 2 \beta \gamma_E \psi_{02} m_s \zeta_1 - 2 \psi_{10} m_s \zeta_1 \nonumber\\
 &-& 10 \beta \psi_{10} m_s \zeta_1 - 5 \psi_{11} m_s \zeta_1 - \beta \psi_{11} m_s \zeta_1 - 10 \gamma_E \psi_{11} m_s \zeta_1 -2 \beta \gamma_E \psi_{11} m_s \zeta_1 - 5 \psi_{12} m_s \zeta_1 - \beta \psi_{12} m_s \zeta_1 \nonumber\\
& -& 10 \gamma_E \psi_{12} m_s \zeta_1 - 2 \beta \gamma_E \psi_{12} m_s \zeta_1 -2 \psi_{21} m_s \zeta_1 - 10 \beta \psi_{21} m_s \zeta_1 + 2 (-2 (\beta -1) (\psi_{10} + \psi_{21}) m_b - (5 + \beta) \big[6 (1 + \gamma_E) \psi_{00} \nonumber\\
&+& 2 \psi_{01} - (5 + 2 \gamma_E) \psi_{02} - (1 +2 \gamma_E) (\psi_{11} + \psi_{12})\Big] m_s) \zeta_2 -4 (5 + \beta) \psi_{00} m_s (\zeta_1 - 2 \zeta_2) \ln(\frac{M^2}{\Lambda^2})\nonumber\\
& -& 2 (5 + \beta) \psi_{02} m_s (\zeta_1 - 2 \zeta_2) \ln(\frac{\Lambda^2}{s}) +2 \Big\{-(5 + \beta) \psi_{02} m_s (\zeta_1 - 2 \zeta_2) \ln(\frac{s}{\Lambda^2}) - m_b ((11 + 13 \beta) \zeta_1 \nonumber\\
&-& 2 (\beta-1 ) \zeta_2) \ln(\frac{s}{m_b^2}) - (5 + \beta) m_s (\zeta_1 -2 \zeta_2) \big[(2 \psi_{00} - \psi_{02} - 2 \psi_{10} + 2 \psi_{21} + \psi_{22}) \ln(\frac{(s-m_b^2 )}{\Lambda^2}) \nonumber\\
&+&\psi_{00} \ln(\frac{(m_b^2 (s-m_b^2 ))}{\Lambda^2 s}) -2 \psi_{02} \ln(\frac{(s (s-m_b^2))}{\Lambda^2 m_b^2})\big]\Big\}\Big)\Big) m_{\pi}^2 + 
 2 m_b^4 (\zeta'_5 - 2 \zeta'_6) \big[\psi_{10} + 5 \beta \psi_{10} - 
 3 (1 + \beta) (\psi_{20} - \psi_{31})\nonumber\\
 & +& (1 + 5 \beta) \ln(\frac{m_b^2}{s})\big] \mu_{\pi}\Bigg]+\frac{(\beta-1)}{384 \pi^2} \mathbb A(u)(u_0) f_{\pi} m_{\pi}^2\Bigg[(\beta-1) (\psi_{10} + \psi_{21}) m_b + (5 + \beta) \psi_{00} m_s \Bigg]\nonumber\\
 &-&\frac{(\beta-1)}{192 \pi^2}m_b^2 f_{\pi}\varphi_{\pi}(u_0) \Bigg[ (\beta -1) (2 \psi_{10} - \psi_{20} + \psi_{31}) m_b + (5 + \beta) (\psi_{20} - \psi_{31}) m_s + 
 2 (\beta -1) m_b \ln(\frac{m_b^2}{s})\Bigg]\nonumber\\
 &-&\frac{(\beta-1)}{576 \pi^2}m_b(-1 + \tilde{\mu}_{\pi}^2) \mu_{\pi} \varphi_{\sigma}(u_0)\Bigg[(5 + b) (\psi_{20} - \psi_{31}) m_b + 2 (\beta -1) (\psi_{10} + \psi_{21}) m_s\Bigg]
+
\frac{\langle\bar{s}s\rangle}{72} (\beta-1) (5 \nonumber\\
&+& \beta) \psi_{00} f_{\pi} \varphi_{\pi}(u_0)
+
\langle g^2G^2\rangle \Bigg\{\frac{1}{576 m_b^4 \pi^2} (\beta-1) (\beta -1) m_s (3 (5 + \beta) \zeta_4 - 4 (1 + 2 \beta) \zeta_1 + (5 + \beta) \zeta_2) f_{\pi} m_{\pi}^2 \Bigg[ 2 \psi_{00} - \psi_{01} \nonumber\\&-& \psi_{02} - 9 \psi_{10} + 6 \gamma_E \psi_{10} + 3 \psi_{21} + \psi_{22} - 6\psi_{10} \ln(\frac{M^2}{\Lambda^2}) + 
 2 (\psi_{00} - \psi_{03} + 3 \psi_{21} + 2 \psi_{22} + \psi_{23}) \ln(\frac{s-m_b^2}{\Lambda^2}) \Bigg]\nonumber\\
 &-&\frac{1}{768 m_b^6 \pi^2} 5 (\beta -1) (5 + \beta) m_s \mathbb A(u)(u_0) f_{\pi}m_{\pi}^2\Bigg[ m_b^2 \Big(6 (2 \gamma_E -3) \psi_{10} + 6 \psi_{21} + 3 \psi_{22} + \psi_{23} - 4 \psi_{-10} + 3 \psi_{-1-1} + \psi_{-1-3}\nonumber\\
 & -& 12 \psi_{10} \ln(\frac{Msq}{\Lambda^2})\Big) + 
 12 \psi_{00} (s-m_b^2) \ln(\frac{s-m_b^2}{\Lambda^2})\Bigg] +\frac{1}{6912 m_b^2 \pi^2}(\beta -1) f_{\pi}\varphi_{\pi}(u_0)\Bigg[ (\beta -1 ) (\psi_{00} - 3 (\psi_{10} + \psi_{21})) m_b \nonumber\\
 & +&  3 (5 + \beta) \Big(3 (-3 + 4 \gamma_E) \psi_{00} - \psi_{01} + 2 \psi_{03} + 5 \psi_{0-1} + 
 37 \psi_{10} - 36 \gamma_E \psi_{10} - \psi_{12} - 3 \psi_{13} - 6 \psi_{21} - 2 \psi_{22}\Big) m_s \nonumber\\
 &-& 6 (5 + \beta) m_s \Big(6 (\psi_{00} - 3 \psi_{10}) \ln(\frac{M^2}{\Lambda^2}) + \big[-19 \psi_{00} - 2 \psi_{03} + 12 \psi_{0-1} + 6 \psi_{21} + 4 \psi_{22} + 2 \psi_{23}\big] \ln(\frac{s-m_b^2}{\Lambda^2}) \nonumber\\
 &+& 3 \psi_{03} \ln(\frac{s (s-mqb^2)}{\Lambda^2 m_b^2)})\Big)\Bigg]+\frac{1}{576 m_b^3 \pi^2} (\beta -1)^2 m_s ( \tilde{\mu}_{\pi}^2 -1 ) \mu_{\pi} \varphi_{\sigma}(u_0)\Bigg[2 \psi_{00} - \psi_{01} - \psi_{02} - 9 \psi_{10} \nonumber\\
 &+& 6 \gamma_E \psi_{10} + 3 \psi_{21} + \psi_{22} - 6 \psi_{10} \ln(\frac{M^2}{\Lambda^2}) + 
 2 (\psi_{00} - \psi_{03} + 3 \psi_{21} + 2 \psi_{22} + \psi_{23}) \ln(\frac{s-m_b^2}{\Lambda^2}) \Bigg]\Bigg\} 
\end{eqnarray}

\begin{eqnarray}
\rho_2(s) &=& \frac{1}{192 m_b \pi^2} m_{\pi}^2\Bigg[ f_{\pi} \Big(m_b^2 \Big(\psi_{10} m_b \big[-2 (1 + \beta (7 + \beta)) \zeta'_3 + 4 (1 + \beta (7 + \beta)) \zeta'_4 - (\beta-1)^2 (\zeta'_7 - 2 \zeta'_8)\big] \nonumber\\
&+& 3 (1 + \beta (4 + \beta)) (\psi_{20} - \psi_{31}) m_b (\zeta'_7 - 2 \zeta'_8) + 
 2 (\beta-1 ) (5 + \beta) \psi_{10} m_s \zeta_B + (5 + \beta) \big[(1 + 5 \beta) (\psi_{20} - \psi_{31}) m_b\nonumber\\
 & +& 2 (\beta-1) \psi_{21} m_s\big] \zeta_B - m_b \big[2 (1 + \beta (7 + \beta)) \zeta'_3 - 4 (1 + \beta (7 + \beta)) \zeta'_4 + (\beta-1)^2 (\zeta'_7 - 2 \zeta'_8)\big] \ln(\frac{m_b^2}{s})\Big) \nonumber\\
 &-& 2 \Big(-(\beta-1) \Big[\big[2 (5 + \beta) (\psi_{01} - 2 \psi_{02}) - (1 + 5 \beta) \psi_{11}\big] m_s (\tilde{\zeta}_1 + \tilde{\zeta}_2) + \psi_{21} \big[(5 + \beta) m_s (\tilde{\zeta}_1 + \tilde{\zeta}_2) + (\beta-1 ) m_b (\tilde{\zeta}_1 + \tilde{\zeta}_2 \nonumber\\
 &+& \tilde{\zeta}_3 - 2\tilde{\zeta}_4 + \tilde{\zeta}_7 - 2 \tilde{\zeta}_8)\big] + \psi_{10} \big[(5 + \beta) (-7 + 6 \gamma_E) m_s (\tilde{\zeta}_1 + \tilde{\zeta}_2) + (\beta-1) m_b (\tilde{\zeta}_1 + \tilde{\zeta}_2 + \tilde{\zeta}_3-2\tilde{\zeta}_4 + \tilde{\zeta}_7 - 2 \tilde{\zeta}_8)\big]\Big] \nonumber\\
 &+& \psi_{00} \Big[3 (\beta-1 ) (-2 + \beta (-2 + \gamma_E) + 5 \gamma_E) m_s (\tilde{\zeta}_1 + \tilde{\zeta}_2) + (5 + \beta) (1 + 5 \beta) m_b (\tilde{\zeta}_1 +\tilde{\zeta}_2 + \tilde{\zeta}_3 -2\tilde{\zeta}_4+ \tilde{\zeta}_7 - 2 \tilde{\zeta}_8)\Big] \nonumber\\
 &-& (\beta-1) (5 + \beta) m_s (\tilde{\zeta}_1 + \tilde{\zeta}_2) \Big[\psi_{00} (M^2-1) \ln(\frac{\Lambda^2}{m_b^2}) + 3 (\psi_{00} - 2 \psi_{10}) \ln(\frac{M^2}{\Lambda^2}) - 2\psi_{02}  \ln(\frac{\Lambda^2}{s}) + \big[-2 \psi_{00} \nonumber\\
 &-& 3 \psi_{02} + 6 \psi_{21} + 3 \psi_{22}\big] \ln(\frac{s-m_b^2}{\Lambda^2}) + 2 \psi_{02} \ln(\frac{s (s-m_b^2)}{\Lambda^2 m_b^2})\Big]\Big) m_{\pi}^2 \Big)-(\beta -1) (\psi_{10} + \psi_{21}) m_b \big[(5 + \beta) m_b \nonumber\\
 &+& (\beta-1 ) m_s\big] (\psi_{72} - 2 \psi_{73}) \mu_{\pi}\Bigg]-\frac{1}{384 \pi^2}((1 + \beta (7 + \beta)) (\psi_{20} - \psi_{31}) m_b^2 f_{\pi} m_{\pi}^2 \mathbb A'(u_0))-\frac{1}{576 \pi^2}(1 + \beta (7 + \beta)) m_b^4\nonumber\\
 &\times& f_{\pi}\varphi_{\pi}'(u_0)\Bigg[6 \psi_{10} - 3 \psi_{20} + \psi_{30} - 2 \psi_{41} - 6 \psi_{00} \ln(\frac{s}{m_b^2}) \Bigg]-\frac{1}{1152 \pi^2}(\beta-1) m_b^2(\tilde{\mu}_{\pi}^2-1 ) \mu_{\pi} \varphi_{\sigma}'(u_0)\Bigg[(5 + b) (2 \psi_{10} \nonumber\\
 &-& \psi_{20} + \psi_{31}) m_b + (\beta-1) (\psi_{20} - \psi_{31}) m_s + 2 (5 + \beta) m_b \ln(\frac{m_b^2}{s})\Bigg]
 +
\frac{\langle\bar{s}s\rangle}{432} \psi_{00} \Bigg[6 (1 + \beta (7 + \beta)) m_s f_{\pi} \varphi_{\pi}'(u_0) \nonumber\\
&+& (\beta-1)^2 (\tilde{\mu}_{\pi}^2 -1 ) \mu_{\pi} \varphi_{\sigma}'(u_0)\Bigg]
+
\langle g^2G^2\rangle \Bigg\{ -\frac{1}{1152 m_b^7 \pi^2}(\beta-1) m_s m_{\pi}^2\Bigg[2 f_{\pi}\Big( (3 (5 + \beta) m_b^4 \zeta_B \Big[2 \psi_{00} - \psi_{01} - \psi_{02} \nonumber\\&-& 9 \psi_{10} + 6 \gamma_E \psi_{10} + 3 \psi_{21} + \psi_{22} - 6 \psi_{10} \ln(\frac{M^2}{\Lambda^2}) + 2 (\psi_{00} - \psi_{03} + 3 \psi_{21} + 2 \psi_{22} + \psi_{23}) \ln(\frac{s-m_b^2}{\Lambda^2})\Big] \nonumber\\
&+& (\tilde{\zeta}_1 +\tilde{\zeta}_2) \Big\{\Big(30 \psi_{00} - 15 \psi_{01} - 15 \psi_{02} - 15 \psi_{0-1} + 12 (29 - 19 \gamma_E) \psi_{10} - 5 \big[-3 + 21 \psi_{21} + 12 \psi_{22} + 5 \psi_{23} - 20 \psi_{-10} \nonumber\\
&+& 5 \psi_{-13} + 15 \psi_{-1-1}\big] +  \beta \big[6 \psi_{00} - 3 \psi_{01} - 3 \psi_{02} - 75 \psi_{0-1} + 12 (19 - 11 \gamma_E) \psi_{10} - 21 \psi_{21} - 12 \psi_{22} - 5 (-15 + \psi_{23} \nonumber\\
&-& 4 \psi_{-10} + \psi_{-13} + 3 \psi_{-1-1})\big]\Big) m_b^2 
+ 12 (19 + 11 \beta) \psi_{10} m_b^2 \ln(\frac{M^2}{\Lambda^2}) + 3 \Big[(-19 + 135 \psi_{00} - 10 \psi_{03} - 6 \psi_{0-1} + 30 \psi_{21} \nonumber\\
&+&
 20 \psi_{22} + 10 \psi_{23} +\beta \big[25 + 27 \psi_{00} - 2 \psi_{03} - 30 \psi_{0-1} + 6 \psi_{21} + 4 \psi_{22} +2 \psi_{23})\big] m_b^2 - 20 (5 + \beta) \psi_{00} s\Big] \ln(\frac{s-m_b^2}{\Lambda^2})\Big\} m_{\pi}^2 \Big)\nonumber\\
 &-&3 (\beta-1) m_b^3 (\zeta_5 - 2 \zeta_6) \Big(2 \psi_{00} - \psi_{01} - \psi_{02} - 9 \psi_{10} + 6 \gamma_E \psi_{10} + 3 \psi_{21} + \psi_{22}- 6 \psi_{10} \ln(\frac{M^2}{\Lambda^2}) 
+ 2 \big[\psi_{00} - \psi_{03} + 3 \psi_{21}\nonumber\\
& +& 2 \psi_{22} + \psi_{23}\big] \ln(\frac{s-m_b^2}{\Lambda^2})\Big) \mu_{\pi}\Bigg] -\frac{1}{3456 \pi^2}(1 + \beta (7 + \beta)) (\psi_{10} + \psi_{21}) f_{\pi} \varphi_{\pi}'(u_0)+\frac{1}{41472 m_b^2 \pi^2}(\beta -1)( \tilde{\mu}_{\pi}^2-1) \mu_{\pi}\nonumber\\
&\times & \varphi_{\sigma}'(u_0)\Bigg[(5 + \beta) (\psi_{00} - 3 (\psi_{10} + \psi_{21})) m_b + 
 3 (\beta-1) \Big[(-1 + 20 \gamma_E) \psi_{00} + \psi_{01} + 2 \psi_{02} - 3 \psi_{03} + 61 \psi_{10} -
\psi_{12} \nonumber\\
&-& 3 \psi_{13} - 60 \gamma_E (\psi_{10} - 2 \psi_{20}) - 181 \psi_{20} + \psi_{31}\Big] m_s+ 
 6 (\beta-1 ) m_s \Big[-10 (\psi_{00} - 3 \psi_{10} + 6 \psi_{20}) \ln(\frac{M^2}{\Lambda^2}) + \big[89 +\psi_{00} \nonumber\\
 &+& \psi_{03}- 135 \psi_{0-1} + 54 \psi_{0-2} + 6 \psi_{31} + 3 \psi_{32} + \psi_{33}\big] \ln(\frac{s-m_b^2}{\Lambda^2})\Big]\Bigg]\Bigg\}
\end{eqnarray}

\begin{eqnarray}
\Gamma_1&=&\langle  \bar{s}s \rangle\Bigg\{ \frac{1}{1728 M^6} (\beta-1)\Bigg[ -\Bigg(6 M^4 \big[4 (5 + \beta) M^2 \zeta_4 - 2 (1 + 5 \beta) M^2 \zeta_1 + m_b m_s (11 \zeta_1 + 13 \beta \zeta_1 + 2 \zeta_2 - 2 \beta \zeta_2)\big]\nonumber\\ 
&+& m_0^2 m_b \big[3 m_b M^2 (-2 (5 + \beta) \zeta_4 + \zeta_1 + 5 \beta \zeta_1) - m_b^2 m_s (11 \zeta_1 + 13 \beta \zeta_1 + 2 \zeta_2 - 2 \beta \zeta_2) + m_s M^2 (11 \zeta_1 + 13 \beta \zeta_1 + 2 \zeta_2\nonumber\\
& -& 2 \beta \zeta_2)\big]\Bigg) f_{\pi} m_{\pi}^2 + (1 + 5 \beta) m_s M^2 (-m_0^2 m_b^2 + 6 M^4) (\zeta'_5 - 
 2 \zeta'_6) \mu_{\pi}\Bigg]-\frac{1}{6912 M^8} (\beta-1)\mathbb A(u)(u_0) f_{\pi} m_{\pi}^2 \Bigg[12 M^4 \Big(m_b^3 (m_s \nonumber\\
 &-& \beta m_s) + 2 (5 + \beta) m_b^2 M^2 + 2 (5 + \beta) M^4\Big) + 
 m_0^2 m_b \Big(2 (\beta -1) m_b^4 m_s + m_b^2 \big[-6 (5 + \beta) m_b + (19 + 5 \beta) m_s\big] M^2 \nonumber\\
 &-& 3 \big[(-3 + \beta) m_b + (5 + 3 \beta) m_s\big] M^4\Big) \Bigg]+\frac{1}{1728 M^4} (\beta -1)f_{\pi} \varphi_{\pi}(u_0)\Bigg[-12 (\beta -1) m_b m_s M^4 + 
 m_0^2 \Big(2 (\beta - 1) m_b^3 m_s \nonumber\\
 &+& 
    3 m_b \big[-2 (5 + \beta) m_b + (5 + 3 \beta) m_s\big] M^2 - 3 (7 + 3 \beta) M^4\Big)\Bigg]+\frac{1}{5184 M^6} (\tilde{\mu}_{\pi}^2-1 ) \mu_{\pi} \varphi_{\sigma}(u_0)\Bigg[ -12 (\beta -1) M^4 ((5 + \beta)\nonumber\\
&\times& m_b^2 m_s - 2 (\beta -1) m_b M^2 + (5 + \beta) m_s M^2) + 
 m_0^2 m_b \Big(2 (\beta -1) (5 + \beta) m_b^3 m_s - (\beta -1) m_b \big[6 (\beta -1) m_b \nonumber\\
 &+& (11 + 7 \beta) m_s\big] M^2 + 24 (1 + \beta + \beta^2) M^4\Big)\Bigg]\Bigg\}
 +
\langle  g^2 G^2 \rangle \Bigg\{ \frac{(\beta-1)}{82944 m_b^2M^6 \pi^2}
\Bigg[ 6 M^2 f_{\pi} \bigg(-(\beta-1) m_b^3 M^2 (2 \zeta_4 + \zeta_1) \nonumber\\
&+& 2 (\beta-1) m_b M^4 (2 \zeta_4 + \zeta_1) -4 m_b^2 m_s M^2 \big[3 (5 + \beta) \zeta_4 - 4 (1 + 2 \beta) \zeta_1 + (5 + \beta) \zeta_2\big] - 
2 m_s M^4 \big[3 (5 + \beta) \zeta_4- 4 (1 + 2 \beta) \zeta_1 \nonumber\\
&+& (5 + \beta) \zeta_2\big]+ 2 m_b^4 m_s \big[2 (5 + \beta) (-1 + 3 \gamma_E) \zeta_4 + (1 + 5 \beta - 
8 (1 + 2 \beta) \gamma_E) \zeta_1 + 2 (5 + \beta) \gamma_E \zeta_2\big] - 
2 m_b^4 m_s \big[3 (5 + \beta) \zeta_4 \nonumber\\
&-& 4 (1 + 2 \beta) \zeta_1 + (5 + \beta) \zeta_2\big] [\ln(\frac{\Lambda^2}{m_b^2}) + 
2 \ln(\frac{M^2}{\Lambda^2})]\bigg) m_{\pi}^2 + 6 (5 + \beta) m_b^2 M^6 (\zeta'_5 - 2 \zeta'_6) \mu_{\pi}\Bigg] +
\frac{1}{27648 m_b^2 M^6 \pi^2}\nonumber\\
&\times &
(\beta-1)  \mathbb A(u)(u_0) f_{\pi}m_{\pi}^2  \Bigg[2 (5 + \beta) (-1 + 3 \gamma_E) m_b^6 m_s - 10 (5 +\beta) m_b^4 m_s M^2 + m_b^2 \big[m_b - \beta m_b - 9 (5 + \beta) m_s \big] M^4 \nonumber\\ 
&+& 2 \big[(\beta-1) m_b - 3 (5 + \beta) m_s\big] M^6 - 
   3 (5 +\beta) m_b^6 m_s [\ln(\frac{\Lambda^2}{m_b^2}) + 2 \ln(\frac{M^2}{\Lambda^2})]\Bigg]+\frac{1}{6912 M^2 \pi^2}(\beta-1)(5 + \beta) m_s f_{\pi} \nonumber\\
&\times &\varphi_{\pi}(u_0)\Bigg[(-2 + 6 \gamma_E) m_b^2 + 6 (-1 + \gamma_E) M^2 - 
 3 (m_b^2 + 2 M^2) \ln(\frac{\Lambda^2}{m_b^2}) - 
 3 (2 m_b^2 + 3 M^2) \ln(\frac{M^2}{\Lambda^2}) \Bigg]\nonumber\\
 &-&\frac{1}{20736 m_b M^4 \pi^2} (\beta-1) (-1 + \tilde{\mu}_{\pi}^2) \mu_{\pi} \varphi_{\sigma}(u_0)\Bigg[2 (\beta-1) (-1 + 3 \gamma_E) m_b^4 m_s - 
 7 (\beta-1) m_b^2 m_s M^2- [(5 + \beta) m_b \nonumber\\
& +& (\beta-1) m_s] M^4 - 3 (\beta-1) m_b^4 m_s [\ln(\frac{\Lambda^2}{m_b^2}) + 2 \ln(\frac{M^2}{\Lambda^2})]\Bigg]\Bigg\}
+
\langle  \bar{s}s \rangle\langle  g^2 G^2 \rangle \Bigg\{\frac{1}{248832 M^{14}}(\beta-1) m_b  f_{\pi} m_{\pi}^2 \mathbb A(u)(u_0)\nonumber\\
&\times &\Bigg[6 M^4 \Big(-(\beta-1) m_b^4 m_s + 
    2 m_b^2 ((5 + \beta) m_b + 3 (\beta-1) m_s) M^2 - 
    2 (2 (5 + \beta) m_b + 3 (\beta-1) m_s) M^4\Big) \nonumber\\
    &+& 
 m_0^2 \Big((\beta-1) m_b^6 m_s - 
    m_b^4 \big[3 (5 +\beta) m_b + 11 (\beta-1) m_s\big] M^2 + 
    6 m_b^2 \big[3 (5 + \beta) m_b + 5 (\beta-1) m_s\big] M^4 - 
    18 \big[(5 + \beta) m_b \nonumber\\
    &+& (\beta-1) m_s\big] M^6\Big) \Bigg]-\frac{1}{62208 M^{10}} (\beta-1) m_b f_{\pi} \varphi_{\pi}(u_0)\Bigg[ 6 M^4 \Big(m_b^2 (m_s -\beta m_s) + 2 (5 + \beta) m_b M^2 + 3 (\beta-1) m_s M^2\Big)\nonumber\\
 &+& m_0^2 \Big((\beta-1) m_b^4 m_s - 
    3 m_b^2 \big[(5 + \beta) m_b + 2 (\beta-1) m_s\big] M^2 + 
    6 \big[(5 + \beta) m_b + (\beta-1) m_s\big] M^4\Big)\Bigg]  
    \nonumber\\
    &-&\frac{1}{186624 M^{12}} (\beta-1) m_b(\tilde{\mu}_{\pi}^2-1) \mu_{\pi} \varphi_{\sigma}(u_0)\Bigg[m_0^2 \big[(5 + \beta) m_b m_s - 3 (\beta-1) M^2\big] (m_b^4 - 6 m_b^2 M^2 + 
    6 M^4) \nonumber\\
    &-& 6 M^4 \Big((5 + \beta) m_b^3 m_s - 2 m_b \big[(-1 + \beta) m_b + (5 + \beta) m_s\big] M^2 + 6 (\beta-1) M^4\Big)\Bigg]\Bigg\} 
\end{eqnarray}

\begin{eqnarray}
\Gamma_2 &=&\frac{1}{96 \pi^2}(\beta-1) (5 + \beta) m_b m_s (\tilde{\zeta}_1 +\tilde{\zeta}_2) f_{\pi}m_{\pi}^4\Bigg[2 \gamma_E - \ln(\frac{\Lambda^2}{m_b^2}) - 2 \ln(\frac{M^2}{\Lambda^2})\Bigg] 
+
\langle  \bar{s}s \rangle\Bigg\{ \frac{1}{1728 M^8}\Bigg[ f_{\pi} m_{\pi}^2\Bigg( M^2 \Big\{6 M^4 \big(m_s M^2 \nonumber\\
&\times & 
\big[-2 (1 + \beta (7 + \beta)) \zeta'_{3} + 4 (1 + \beta (7 + \beta)) \zeta'_{4} - (\beta-1)^2 (\zeta'_{7} - 2 \zeta'_{8})\big] + 2 (5 + \beta) (m_b^2 (m_s + 5 \beta m_s) - 2 (\beta-1) m_b M^2 \nonumber\\
&+& (1 + 5 \beta) m_s M^2) \zeta_B \big) + m_0^2 m_b \Big[-2 (5 + \beta) (1 + 5 \beta) m_b^3 m_s \zeta_B + 6 (\beta-1) (5 + \beta) m_b^2 M^2 \zeta_B - 18 (\beta^2-1) M^4 \zeta_B \nonumber\\
&+& m_b m_s M^2 \big[2 (1 + \beta (7 + \beta)) \zeta'_3 - 4 (1 + \beta (7 + \beta)) \zeta'_{4} + (\beta-1)^2 (\zeta'_{7} - 2 \zeta'_{8} + 3 \zeta_B)\big]\Big]\Big\} - 2 (\beta-1) m_b \Big[6 M^4 \big[2 (1 + 5 \beta)\nonumber\\
&\times & M^2 (\tilde{\zeta}_1 + \tilde{\zeta}_2) - (\beta-1) m_b m_s (\tilde{\zeta}_1 + \tilde{\zeta}_2 + \tilde{\zeta}_3-2\tilde{\zeta}_4 + \tilde{\zeta}_7 - 2 \tilde{\zeta}_8)\big] + m_0^2 \big(-3 (1 + 5 \beta) m_b^2 M^2 (\tilde{\zeta}_1 + \tilde{\zeta}_2) + 3 (1 + 5 \beta) M^4 (\tilde{\zeta}_1 \nonumber\\
&+& \tilde{\zeta}_2) + (\beta-1) m_b^3 m_s (\tilde{\zeta}_1 + \tilde{\zeta}_2 +\tilde{\zeta}_3-2\tilde{\zeta}_4 + \tilde{\zeta}_7 - 2 \tilde{\zeta}_8)- 2 (\beta-1) m_b m_s M^2 (\tilde{\zeta}_1 + \tilde{\zeta}_2 + \tilde{\zeta}_3-2\tilde{\zeta}_4 + \tilde{\zeta}_7 - 2 \tilde{\zeta}_8)\big)\Big] m_{\pi}^2\Bigg) \nonumber\\
&-& 3 (\beta-1)^2 M^4 \mu_{\pi} \Big((m_0^2 m_b^2 - 4 M^4) (\zeta_5 - 2 \zeta_6) m_{\pi}^2 + m_0^2 M^4 (2v- 1) \varphi_P(u_0)\Big)\Bigg]+
\frac{1}{6912 M^6}m_s f_{\pi} m_{\pi}^2 \mathbb A'(u_0)\Bigg[-24 (1 \nonumber\\
&+& \beta (7 + \beta)) M^4 (m_b^2 + M^2) + 
 m_0^2 m_b^2 \big[4 (1 + \beta (7 + \beta)) m_b^2 + (7 + \beta (4 + 7 \beta)) M^2\big]\Bigg]-\frac{1}{1728 M^2}m_0^2 m_s f_{\pi} \varphi'_{\pi}(u_0)\Bigg[ 4 (1 \nonumber\\
 &+& \beta (7 + \beta)) m_b^2 + (11 + \beta (32 + 11 \beta)) M^2\Bigg]+\frac{1}{10368 M^4}(\beta
 -1) (  \tilde{\mu}_{\pi}^2-1) \mu_{\pi} \varphi'_{\sigma}(u_0)\Bigg[ -12 (5 + \beta) m_b m_s M^4 + 
 m_0^2 \Big(2 (5 \nonumber\\
 &+& \beta) m_b^3 m_s + 3 m_b (-2 (\beta-1) m_b + (\beta-3) m_s) M^2 - 3 (3 + 5 \beta) M^4\Big)\Bigg]\Bigg\}
 +
\langle  g^2 G^2 \rangle \Bigg\{ \frac{1}{13824 m_b^3 M^6\pi^2}m_{\pi}^2 \Bigg[ f_{\pi} \Bigg(m_b^2 M^2\nonumber\\
&\times & \Big(m_b M^4 \big[2 (1 + \beta (7 + \beta)) \zeta'_3 - 4 (1 + \beta (7 + \beta)) \zeta'4 - (5 + \beta) (1 + 5 \beta) (\zeta'_7 - 2 \zeta'_8)\big] + 2 (5 + \beta) \big[2 (\beta-1) (3 \gamma_E-1) m_b^4 m_s \nonumber\\
&-& 7 (\beta-1) m_b^2 m_s M^2 - (m_b + 5 \beta m_b + (\beta-1) m_s) M^4\big] \zeta_B 
- 6 (\beta-1) (5 + \beta) m_b^4 m_s \zeta_B \big[\ln(\frac{\Lambda^2}{m_b^2}) + 2 \ln(\frac{Msq}{\Lambda^2})\big]\Big) \nonumber\\
&+& 2 \Big(2 (\beta-1) (-1 - 2 \gamma_E + \beta (14 \gamma_E -5 )) m_b^6 m_s (\tilde{\zeta}_1 + \tilde{\zeta}_2)- 8 (\beta -1) (-1 + 4 \beta + 3 (\beta-1) \gamma_E) m_b^4 m_s M^2 (\tilde{\zeta}_1 + \tilde{\zeta}_2) \nonumber\\&+& 
12 (\beta-1)^2 m_b^2 m_s M^4 (\tilde{\zeta}_1 + \tilde{\zeta}_2) - (\beta -1) (5 + \beta) m_s M^6 (\tilde{\zeta}_1 + \tilde{\zeta}_2) + (5 + \beta) (1 +5 \beta) m_b^5 M^2 (\tilde{\zeta}_1 + \tilde{\zeta}_2 + \tilde{\zeta}_3-2\tilde{\zeta}_4 \nonumber\\
&+& \tilde{\zeta}_7 - 2 \tilde{\zeta}_8) 
+ (\beta -1) m_b^4 m_s (\tilde{\zeta}_1 + 
\tilde{\zeta}_2) \Big[\big[(2 - 14 \beta) m_b^2 + 9 (3\beta -1) M^2\big] \ln(\frac{\Lambda^2}{m_b^2}) + \big[(4 - 28 \beta) m_b^2 + 3 (-7 + 13 \beta) M^2\big] \nonumber\\
&\times& \ln(\frac{Msq}{\Lambda^2})\Big]\Big) m_{\pi}^2\Bigg) 
- (\beta -1) m_b M^2 (\zeta_5 - 2 \zeta_6) \Bigg(2 (\beta -1) (3 \gamma_E -1) m_b^4 m_s - m_b^2 ((5 + \beta) m_b + 6 (\beta -1) m_s) M^2\nonumber\\ &+& (2 (5 + \beta) m_b 
- 3 (\beta -1 ) m_s) M^4 - 3 (\beta -1 ) m_b^4 m_s \big[\ln(\frac{\Lambda^2}{m_b^2}) + 2 \ln(\frac{M^2}{\Lambda^2})\big]\Bigg) \mu_{\pi}\Bigg]+\frac{1}{13824 \pi^2}(1 + \beta (7 + \beta)) f_{\pi} m_{\pi}^2 \mathbb A'(u_0)\nonumber\\
&-&\frac{1}{41472 M^2 \pi^2} (\beta -1)^2 m_s (\tilde{\mu}_{\pi}^2 -1) \mu_{\pi} \varphi'_{\sigma}(u_0)\Bigg[ (6 \gamma_E - 2 ) m_b^2 + 6 (\gamma_E -1) M^2 - 3 (m_b^2 + 2 M^2) \ln(\frac{\Lambda^2}{m_b^2}) \nonumber\\
&-& 3 (2 m_b^2 + 3 M^2) \ln(\frac{M^2}{\Lambda^2})\Bigg] \Bigg\}
+
\langle  \bar{s}s \rangle\langle  g^2 G^2 \rangle \Bigg\{\frac{1}{62208 M^{12}}(5 + \beta) m_b \zeta_B f_{\pi} m_{\pi}^2 \Bigg[m_0^2 \big[m_b (m_s + 5 \beta m_s) - 3 (\beta-1) M^2\big] (m_b^4 \nonumber\\
&-& 6 m_b^2 M^2 + 6 M^4) + 6 M^4 \Big(-(1 + 5 \beta) m_b^3 m_s + 2 m_b \big[(\beta -1) m_b + m_s + 5 \beta m_s\big] M^2 - 6 (\beta -1) M^4\Big)\Bigg]\nonumber\\
&-&\frac{1}{124416 M^{12}}(1 + \beta (7 + \beta)) m_b^2 m_s f_{\pi} m_{\pi}^2 \mathbb A'(u_0) \Bigg[6 M^4 (-m_b^2 + 2 M^2) + m_0^2 (m_b^4 - 6 m_b^2 M^2 + 6 M^4)\Bigg]\nonumber\\
&+& \frac{1}{31104 M^8} (1 + \beta (7 + \beta)) m_b^2 m_s f_{\pi} \varphi'_{\pi}(u_0)(m_0^2 (m_b^2 - 2 M^2) - 6 M^4) -\frac{1}{373248 M^{10}} (\beta -1) m_b(\tilde{\mu}_{\pi}^2-1) \mu_{\pi}\nonumber\\
&\times & \varphi'_{\sigma}(u_0) \Bigg[ 6 M^4 \big[-(5 + \beta) m_b^2 m_s + 2 (\beta -1) m_b M^2 + 
 3 (5 + \beta) m_s M^2\big] + m_0^2 \Big((5 + \beta) m_b^4 m_s \nonumber\\
 &-&3 m_b^2 \big[(\beta -1) m_b + 2 (5 + \beta) m_s\big] M^2 +  6 \big[(\beta -1) m_b + (5 + \beta) m_s\big] M^4\Big)\Bigg]\Bigg\} 
\end{eqnarray}

where the functions $\zeta_{j}$, $\zeta'_{j}$, $\tilde{\zeta}_{j}$ and $\zeta_{B}$ are defined as
\begin{eqnarray}\label{etalar}
\zeta_{j} &=& \int {\cal D}\alpha_i \int_0^1 dv f_{j}(\alpha_i)
\delta(k(\alpha_{ q} +v \alpha_g) -  u_0),
\nonumber \\
\zeta'_{j} &=& \int {\cal D}\alpha_i \int_0^1 dv f_{j}(\alpha_i)
\delta'(k(\alpha_{ q} +v \alpha_g) -  u_0),
\nonumber \\
\tilde{\zeta}_{j} &=& \int {\cal D}\alpha_i \int_0^1 dv f_{j}(\alpha_i)
\theta(k(\alpha_{ q} +v \alpha_g) -  u_0),
\nonumber \\
\zeta_{B} &=& \int_{u_0}^{1} du' \mathbb B(u'),
\nonumber \\
\psi_{nm}&=&\frac{{( {s-m_{Q}}^2 )
}^n}{s^m{(m_{Q}^{2})}^{n-m}},\nonumber \\
\end{eqnarray}
 with  $f_{1}(\alpha_i)={\cal V_{\parallel}}(\alpha_i)$, $f_{2}(\alpha_i)={\cal V_{\perp}}(\alpha_i)$,
 $f_{3}(\alpha_i)={\cal A_{\parallel}}(\alpha_i)$, $f_{4}(\alpha_i)={\cal
 T}(\alpha_i)$, $f_{5}(\alpha_i)=v{\cal T}(\alpha_i)$ being the pion distribution amplitudes whose explicit forms can be found in Refs.~\cite{Belyaev:1994zk,Ball:2004ye,Ball:2004hn}. Note that, $u_{0}=\frac{M_{1}^{2}}{M_{1}^{2}+M_{2}^{2}}$  and due to the close masses of initial and final baryons this expression becomes,
 $u_{0} =\frac{1}{2}$ with the usage of $ M_{1}^{2} = M_{2}^{2} $. In the above results $\mu_{\pi} = f_{\pi} \frac{m_{\pi}^2}{m_{u} + m_{d}},$ $\tilde
\mu_{\pi} = \frac{{m_{u} + m_{d}}}{m_{\pi}}$, ${\cal D} \alpha =
 d \alpha_{\bar q}  d \alpha_q  d \alpha_g
\delta(1-\alpha_{\bar q}-\alpha_q-\alpha_g)$ and the
$\varphi_{\pi}(u),$ $\mathbb A(u),$ $\mathbb B(u),$ $\varphi_P(u),$
$\varphi_\sigma(u),$ ${\cal T}(\alpha_i),$ ${\cal
A}_\perp(\alpha_i),$ ${\cal A}_\parallel(\alpha_i),$ ${\cal
V}_\perp(\alpha_i)$ and ${\cal V}_\parallel(\alpha_i)$ are functions
of definite twist which are given as \cite{Belyaev:1994zk,Ball:2004ye,Ball:2004hn}
\begin{eqnarray}
\varphi_{\pi}(u) &=& 6 u \bar u \left( 1 + a_1^{\pi} C_1(2 u -1) +
a_2^{\pi} C_2^{3 \over 2}(2 u - 1) \right),
\nonumber \\
{\cal T}(\alpha_i) &=& 360 \eta_3 \alpha_{\bar q} \alpha_q
\alpha_g^2 \left( 1 + w_3 \frac12 (7 \alpha_g-3) \right),
\nonumber \\
\varphi_P(u) &=& 1 + \left( 30 \eta_3 - \frac{5}{2} \mu_{\pi}^2 \right)
C_2^{1 \over 2}(2 u - 1)
\nonumber \\
&+& \left( -3 \eta_3 w_3  - \frac{27}{20} \mu_{\pi}^2 -
\frac{81}{10} \mu_{\pi}^2 a_2^{\pi} \right) C_4^{1\over2}(2u-1),
\nonumber \\
\varphi_\sigma(u) &=& 6 u \bar u \left[ 1 + \left(5 \eta_3 - \frac12
\eta_3 w_3 - \frac{7}{20}  \mu_{\pi}^2 - \frac{3}{5} \mu_{\pi}^2
a_2^{\pi} \right) C_2^{3\over2}(2u-1) \right],
\nonumber \\
{\cal V}_\parallel(\alpha_i) &=& 120 \alpha_q \alpha_{\bar q}
\alpha_g \left( v_{00} + v_{10} (3 \alpha_g -1) \right),
\nonumber \\
{\cal A}_\parallel(\alpha_i) &=& 120 \alpha_q \alpha_{\bar q}
\alpha_g \left( 0 + a_{10} (\alpha_q - \alpha_{\bar q}) \right),
\nonumber\\
{\cal V}_\perp (\alpha_i) &=& - 30 \alpha_g^2\left[
h_{00}(1-\alpha_g) + h_{01} (\alpha_g(1-\alpha_g)- 6 \alpha_q
\alpha_{\bar q}) +
    h_{10}(\alpha_g(1-\alpha_g) - \frac32 (\alpha_{\bar q}^2+ \alpha_q^2)) \right],
\nonumber\\
{\cal A}_\perp (\alpha_i) &=& 30 \alpha_g^2(\alpha_{\bar q} -
\alpha_q) \left[ h_{00} + h_{01} \alpha_g + \frac12 h_{10}(5
\alpha_g-3) \right],
\nonumber \\
\mathbb B(u)&=& g_{\pi}(u) - \phi_{\pi}(u),
\nonumber \\
g_{\pi}(u) &=& g_0 C_0^{\frac12}(2 u - 1) + g_2 C_2^{\frac12}(2 u -
1) + g_4 C_4^{\frac12}(2 u - 1),
\nonumber \\
{\mathbb A}(u) &=& 6 u \bar u \left[\frac{16}{15} + \frac{24}{35}
a_2^{\pi}+ 20 \eta_3 + \frac{20}{9} \eta_4 +
    \left( - \frac{1}{15}+ \frac{1}{16}- \frac{7}{27}\eta_3 w_3 - \frac{10}{27} \eta_4 \right) C_2^{3 \over 2}(2 u - 1)
    \right. \nonumber \\
    &+& \left. \left( - \frac{11}{210}a_2^{\pi} - \frac{4}{135} \eta_3w_3 \right)C_4^{3 \over 2}(2 u - 1)\right]
\nonumber \\
&+& \left( -\frac{18}{5} a_2^{\pi} + 21 \eta_4 w_4 \right)\left[ 2
u^3 (10 - 15 u + 6 u^2) \ln u \right. \nonumber\\ &+& \left. 2 \bar
u^3 (10 - 15 \bar u + 6 \bar u ^2) \ln\bar u + u \bar u (2 + 13 u
\bar u) \right] \label{wavefns},
\end{eqnarray}
where $C_n^k(x)$ are the Gegenbauer polynomials,
\begin{eqnarray}
h_{00}&=& v_{00} = - \frac13\eta_4,
\nonumber \\
h_{01} &=& \frac74  \eta_4 w_4  - \frac{3}{20} a_2^{\pi},
\nonumber \\
h_{10} &=& \frac74 \eta_4 w_4 + \frac{3}{20} a_2^{\pi},
\nonumber \\
a_{10} &=& \frac{21}{8} \eta_4 w_4 - \frac{9}{20} a_2^{\pi},
\nonumber \\
v_{10} &=& \frac{21}{8} \eta_4 w_4,
\nonumber \\
g_0 &=& 1,
\nonumber \\
g_2 &=& 1 + \frac{18}{7} a_2^{\pi} + 60 \eta_3  + \frac{20}{3}
\eta_4,
\nonumber \\
g_4 &=&  - \frac{9}{28} a_2^{\pi} - 6 \eta_3 w_3 \label{param0}.
\end{eqnarray}
 Eqs.~(\ref{wavefns}) and (\ref{param0}) contain some constants which were
calculated using QCD sum rules at the renormalization scale $\mu=1$~GeV$^{2}$ \cite{Ball:2006wn,Belyaev:1994zk,Ball:2004ye,Ball:2004hn,R23,ek1,ek2,ek3} and are given as $a_{1}^{\pi} = 0$,
$a_{2}^{\pi} = 0.44$, $\eta_{3} =0.015$, $\eta_{4}=10$, $w_{3} = -3$
and $ w_{4}= 0.2$.

%%%%%%%%%%%%%%%%%%%%%%%%%%%%%%%%%%%%%%%%%%%%%%%%%%%%%%%%%%%%%%%%%%%%%%%%%%%%%%%%%%%%%%%%%%%%%%%%%%%%%%%%%%%%

\end{document}